\documentclass[a4paper,fleqn,usenatbib]{mnras}

\usepackage{newtxtext,newtxmath}

\usepackage[T1]{fontenc}
\usepackage{ae,aecompl}

\usepackage{graphicx}	
\usepackage{amsmath}	
\usepackage{threeparttable}

\title[Photochemical hazes in hot Jupiter atmospheres]{3D simulations of photochemical hazes in the atmosphere of hot Jupiter HD~189733b}

\author[M. E. Steinrueck et al.]{
Maria E. Steinrueck$^{1}$\thanks{E-mail: msteinru@lpl.arizona.edu},
Adam P. Showman\footnote{Deceased}$^{1}$,
Panayotis Lavvas,$^{2}$,
Tommi Koskinen$^{1}$,
\newauthor Xianyu Tan$^{3}$,
and Xi Zhang$^{4}$
\\
$^{1}$Lunar and Planetary Laboratory, University of Arizona, Tucson, AZ, 85721, USA\\
$^{2}$Groupe de Spectrom\'etrie Moleculaire et Atmosph\'erique, Universit\'e de Reims Champagne-Ardenne, Reims, France\\
$^{3}$Atmospheric, Ocean, and Planetary Physics, Department of Physics, Oxford University, OX1 3PU, UK\\
$^{4}$Department of Earth and Planetary Sciences, University of California, Santa Cruz, CA 95064, USA
}

\date{Accepted 2021 April 5. Received 2021 March 12; in original form 2020 November 27 \\
This is a pre-copyedited, author-produced PDF of an article accepted for publication in Monthly Notices of the Royal Astronomical Society following peer review. The version of record (Monthly Notices of the Royal Astronomical Society, Volume 504, Issue 2, June 2021, Pages 2783–2799, doi:10.1093/mnras/stab1053) is available online at: \url{https://academic.oup.com/mnras/article-abstract/504/2/2783/6232173}. This version incorporates the corrected versions of equations (7) and (8), as described in the correction published in July 2024, available at \url{https://academic.oup.com/mnras/article/531/3/3444/7691150}.}

\pubyear{2021}

\begin{document}
\label{firstpage}
\pagerange{\pageref{firstpage}--\pageref{lastpage}}
\maketitle

\begin{abstract}
Photochemical hazes have been suggested as candidate for the high-altitude aerosols observed in the transmission spectra of many hot Jupiters. We present 3D simulations of the hot Jupiter HD 189733b to study how photochemical hazes are transported by atmospheric circulation. The model includes spherical, constant-size haze particles that gravitationally settle and are transported by the winds as passive tracers, with particle radii ranging from 1 nm to 1 $\mu$m. We identify two general types of haze distribution based on particle size: In the small-particle regime ($<$30 nm), gravitational settling is unimportant, and hazes accumulate in two large mid-latitude vortices centered on the night side that extend across the morning terminator. Therefore, small hazes are more concentrated at the morning terminator than at the evening terminator. In the large-particle regime ($>$30 nm), hazes settle out quickly on the nightside, resulting in more hazes at the evening terminator. For small particles, terminator differences in haze mass mixing ratio and temperature considered individually can result in significant differences in the transit spectra of the terminators. When combining both effects for HD189733b, however, they largely cancel out each other, resulting in very small terminator differences in the spectra. Transit spectra based on the GCM-derived haze distribution fail to reproduce the steep spectral slope at short wavelengths in the current transit observations of HD 189733b. Enhanced sub-grid scale mixing and/or optical properties of hazes differing from soot can explain the mismatch between the model and observations, although uncertainties in temperature and star spots may also contribute to the spectral slope.
\end{abstract}




\section{Introduction}
Hot Jupiter HD~189733b is one of the best-studied exoplanets to date. Its transmission spectrum exhibits a particularly strong short-wavelength slope \citep{PontEtAl2008,SingEtAl2011,PontEtAl2013}. In combination with the non-detection of the wings of the sodium line \citep{HuitsonEtAl2012} and the low amplitude of the water feature \citep{SingEtAl2009,GibsonEtAl2012HD189, McCulloughEtAl2014}, this has been interpreted as evidence for an extended layer of high-altitude aerosols \citep{PontEtAl2013,SingEtAl2016}. \citet{McCulloughEtAl2014} pointed out that a significant fraction of the slope could also be caused by unocculted star spots, in which case the small amplitude of the water feature could be caused by a clear atmosphere with a  subsolar water abundance. However, without invoking high-altitude aerosols, the muted wings of the sodium feature remain difficult to explain. Furthermore, strong short-wavelength slopes have been observed in the transmission spectra of a number of hot Jupiters around different stellar types and stellar activity levels \citep[e.g.,][]{NikolovEtAl2015WASP-6b,SingEtAl2016,WongEtAl2020HAT-P-12b,SpakeEtAl2020WASP-127b}.
\citet{SingEtAl2016} examined the correlation between the spectral slope from the near-infrared to mid-infrared and the amplitude of the water feature in a sample of ten hot Jupiters. They found that the observed correlation is consistent with aerosols and inconsistent with highly sub-solar water abundances. These observations suggest that aerosols contribute significantly to the transmission spectra of many hot Jupiters, including HD~189733b.

 The origin of the aerosols is unclear.
 They could either form as the result of photochemical reactions initiated by the UV radiation of the host star at high altitudes (photochemical hazes) or through condensation of gas phase species as air is mixed towards regions with lower temperatures (condensate clouds).
The photochemical haze scenario so far has received less attention from modelers compared to condensate clouds. Yet, photochemical hazes are found in the atmospheres of all giant planets in the Solar System and are also expected to form on short-period extrasolar giant planets. Laboratory experiments show that photochemical hazes readily form over a broad range of conditions expected for short-period exoplanets \citep{HorstEtAl2018,HeEtAl2018,HeEtAl2020} and can form in hydrogen-dominated atmospheres at temperatures as high as 1,500 K \citep{FleuryEtAl2019}, though the latter result strongly depends on the C/O ratio \citep{FleuryEtAl2020}.

Using a combination of a 1D photochemistry-thermochemisty-transport model and a haze microphysics model, \citet{LavvasKoskinen2017} found that hydrocarbon hazes with soot-like properties can explain the spectrum of HD~189733b. \citet{GaoEtAl2020} also showed that photochemical hazes are expected to be the dominant opacity source for hot Jupiters with equilibrium temperatures below 950 K. Haze production rates in their model were based on methane photolysis rates derived using the equilibrium chemistry abundance of methane at low pressures and did not take into account transport-induced quenching. This procedure may underestimate the haze production rate for some hot Jupiters. \citet{LavvasKoskinen2017} found larger haze production rates in comparison based on their more detailed photochemical model. Therefore, haze opacity could be important for hot Jupiters with significantly higher equilibrium temperatures than the 950 K limit inferred by \citet{GaoEtAl2020}. \citet{HellingEtAl2020WASP-43b} found that hydrocarbon hazes even form on the dayside of WASP-43b, which has a zero-albedo equilibrium temperature of 1,400 K.
Despite the clear relevance of photochemical hazes, no studies to date have explored how atmospheric circulation shapes the three-dimensional distribution of photochemical hazes. 

Hazes form predominantly on the dayside but transmission spectra probe the terminator. When using 1D models to interpret transmission spectra, one has to make assumptions about how the haze distribution obtained using dayside-average conditions (typically used in haze microphysics models) relates to the haze distribution at the terminator. Typically, for this purpose, it is assumed that hazes are distributed homogeneously around the planet. Without testing this assumption with 3D general circulation models (GCMs), it is, however, unclear if that assumption is justified. Studies of condensate clouds show that the combination of atmospheric circulation and gravitational settling can produce significant horizontal abundance variations \citep{ParmentierEtAl2013,CharnayEtAl2015a}. More complex models of condensate clouds also point towards the importance of horizontal mixing \citep{LeeEtAl2016,LinesEtAl2018}.

Further, one-dimensional studies treat vertical mixing in a highly simplified way by assuming that it is a purely diffusive process. The strength of vertical mixing then becomes a free parameter, the eddy diffusion coefficient ($K_{zz}$). A common method for estimating the eddy diffusion coefficient is to multiply the root-mean-square of the vertical velocity (taken from a general circulation model) with the atmospheric scale height. \citet{ParmentierEtAl2013} demonstrated that the $K_{zz}$ derived using the distribution of cloud particles in a GCM is significantly lower than the $K_{zz}$ from the root-mean-square velocity.
\citet{ZhangShowman2018FastRotators,ZhangShowman2018TidallyLocked}  further showed that when simulating a gas-phase chemical species with a limited chemical lifetime in two and three dimensions, non-diffusive effects can be significant or even dominant when the species have a long chemical lifetime or when there are horizontal variations in the equilibrium abundance. This is particularly relevant for photochemical species, which are produced on the dayside only.

Understanding the extent of which an inhomogeneous haze coverage can be expected is crucial for interpreting observations correctly and for planning future observations. For example, \citet{LineParmentier2016} demonstrated that partial cloud coverage can lead to bias in the interpretation of transmission spectra. \citet{KemptonEtAl2017} suggested that measuring differences between the morning terminator (leading limb in transit) and evening terminator (trailing limb in transit) through ingress and egress spectroscopy could help distinguish between condensate clouds and photochemical hazes. Their argument was based on the assumption that photochemical hazes would predominantly be found at the evening terminator, because they would be carried eastward from the dayside by the equatorial jet and settle out on the nightside. Condensate clouds, in contrast, would be predominantly found at the morning terminator, which is colder. Follow-up studies show that for condensate clouds, the picture might be more complicated. Even though the total cloud mass is larger at the morning terminator, the hotter temperature profile at the evening terminator can result in condensate clouds forming at lower pressures. This can lead to a larger transit radius at the evening terminator compared to the morning terminator at short wavelengths \citep{PowellEtAl2019TransitSignatures}.  There remains significant interest in investigating terminator differences observationally. A theoretical study of the 3D distribution of photochemical hazes is required to complement these efforts.

The aim of this study is thus to explore how photochemical hazes are mixed globally using GCM simulations of the hot Jupiter HD~189733b. Our model assumes a haze source at low pressures centered on the dayside. The spherical, constant-size haze particles are advected by atmospheric circulation and are subject to gravitational settling. Once hazes reach higher pressures, they are expected to either thermally decompose because of higher temperatures or be removed from the population of `pure' hazes because cloud species condense on them. To represent these processes in a simple fashion, a haze sink is included for pressures higher than 100~mbar. The model is described in more detail in Section \ref{sec:methods}. Section \ref{sec:results} summarizes the results from the GCM, including an overview of the atmospheric circulation and a detailed description of the haze distribution in the two regimes we identify: small particles (<30 nm) and large particles (>30 nm). Section \ref{sec:eddydiffusioncoefficient} examines how the globally averaged haze mixing ratio profile compares to 1D models and infers an effective eddy diffusion coefficient $K_{zz}$. In Section \ref{sec:spectra}, we explore the implications for transit spectra. Section \ref{sec:extrakzz} briefly investigates the possibility that sub-grid-scale turbulence not captured by the GCM results in stronger vertical mixing. Finally, Section \ref{sec:discussion} discusses the limitations of our model and future directions and Section \ref{sec:conclusions} summarizes our conclusions.

\section{Methods}
\label{sec:methods}
We use the MITgcm \citep{AdcroftEtAl2004} to simulate the atmosphere of HD~189733b. This model has been successfully applied to a wide range of exoplanets, including hot Jupiters \citep[e.g.,][]{ShowmanEtAl2009,LiuShowman2013,ParmentierEtAl2013,KatariaEtAl2016,SteinrueckEtAl2019}, highly eccentric hot Jupiters \citep[e.g.,][]{KatariaEtAl2013,LewisEtAl2014}, 
 warm Jupiters \citep{ShowmanEtAl2015}, mini-Neptunes \citep{KatariaEtAl2014,ZhangShowman2017} and terrestrial exoplanets \citep{CaroneEtAl2014}. The radiative transfer in the GCM is calculated using a double-gray two-stream solver \citep{KyllingEtAl1995}  which has previously been used in conjunction with the MITgcm to simulate the atmospheres of hot Jupiters \citep{KomacekEtAl2017,KomacekEtAl2019} and ultra-hot Jupiters \citep{TanKomacek2019}.

\begin{table}	
	\centering	
	\caption{Model parameters}
	\begin{threeparttable}[b]

	\label{tab:modelparameters}
	\begin{tabular}{lrr} 
		\hline
		Parameter & Value & Units\\
		\hline
		Radius \tnote{1}& $1.13$ & $R_J$ \\ 
		Gravity \tnote{1}& 21.93 & m s$^{-2}$ \\ 
		Rotation period \tnote{1} & 2.21857567  & d \\
		Semimajor axis\tnote{2} & 0.03142 & AU \\
		Specific heat capacity & $1.3\cdot 10^4$ & J kg$^{-1}$ K$^{-1}$ \\
		Specific gas constant & 3714 & J kg$^{-1}$ K$^{-1}$ \\
		Interior flux\tnote{3} & 851 & W m$^{-2}$ \\ 
		Horizontal resolution & C32\tnote{a} &  \\
		Vertical resolution & 60 & layers \\
		Lower pressure boundary & $1.75 \cdot 10^{-7}$ & bar \\
		Upper pressure boundary & 200 & bar \\
		Hydrodynamic time step & 25 & s \\
		Radiative time step & 50 & s \\
		\hline
	\end{tabular}
	\begin{tablenotes}
	\item[1] \citet{StassunEtAl2017}
\item[2] \citet{SouthworthEtAl2010III}
\item[3] corresponding to an internal temperature of 350K, based on \citet{ThorngrenEtAl2019InternalTemperature}
\item[a] equivalent to a resolution of 128x64 on a longitude-latitude grid
	\end{tablenotes}
	\end{threeparttable}
\end{table}

\subsection{Dynamics}
\begin{figure*}
\includegraphics[width=\textwidth]{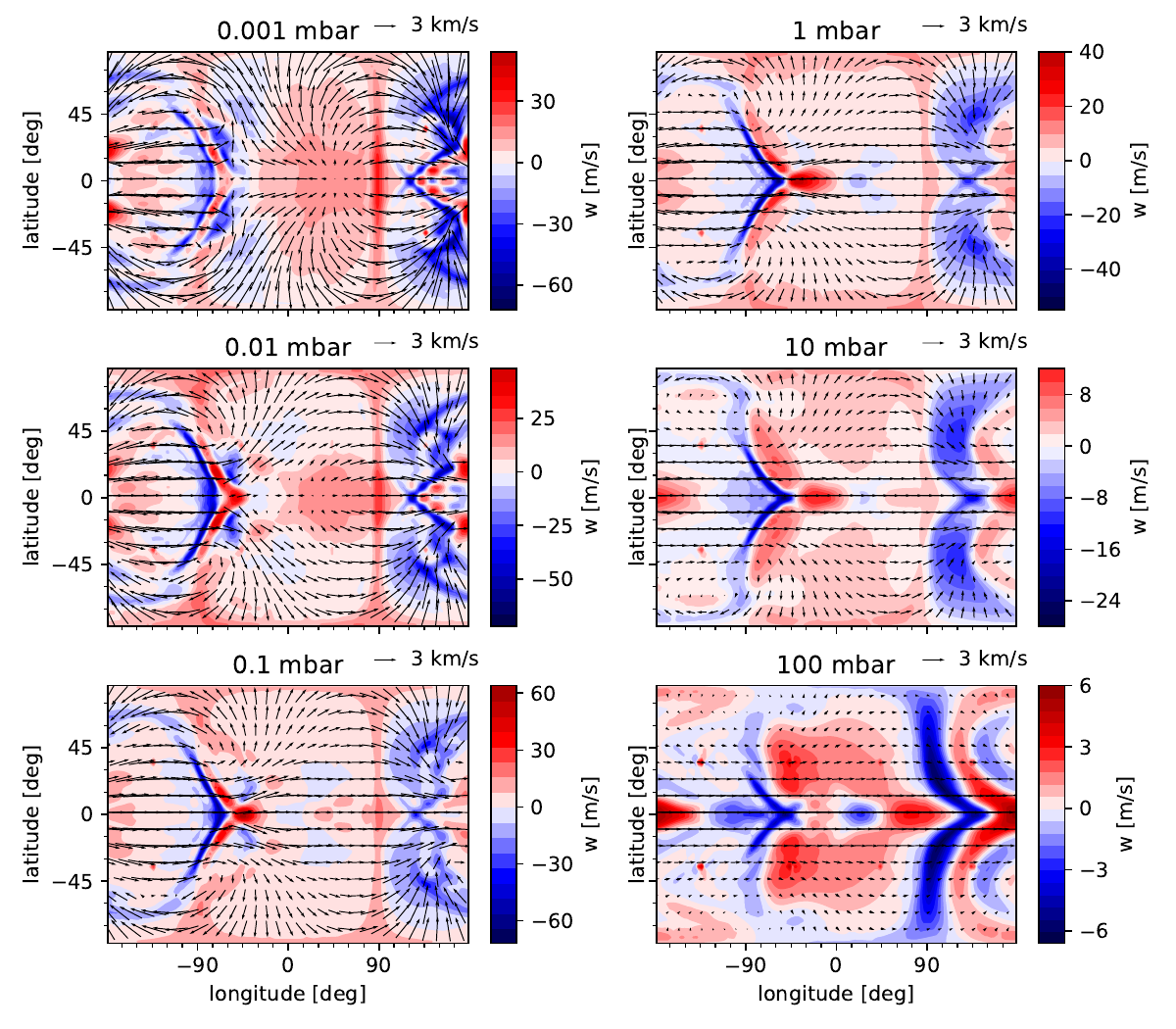}
\caption{Horizontal (arrows) and vertical (colorscale) velocities in our simulation on 6 different isobars. Velocities have been averaged over the last 100 days of the simulation. The substellar point is located at the center of the panel. The morning terminator (leading limb) and evening terminator (trailing limb) are located at a longitude of -90$^\circ$ and 90$^\circ$, respectively. Note that the color scale varies between panels while the arrows remain at the same scale throughout all panels.}
\label{fig:velocities}
\end{figure*}
The dynamical core of the MITgcm \citep{AdcroftEtAl2004} solves the three-dimensional primitive equations on a cubed-sphere grid. The primitive equations describe atmospheric flow well for shallow and stably stratified atmospheres, conditions fulfilled for hot Jupiters \citep[e.g.,][]{ShowmanEtAl2008,MayneEtAl2014}.
Sources of small-scale numerical instability are controlled by applying a fourth-order Shapiro filter \citep{Shapiro1970}.
 To stabilize the flow in the deep atmosphere, a pressure-dependent linear drag $-k_v \textbf{v}$ is applied at pressures $P>10$ bar. The pressure dependence of the drag coefficient $k_v$ takes the same form as in \citet{LiuShowman2013}, with $k_v=k_F (p - p_{\textrm{drag,top}})/(p_{\textrm{bottom}}-p_{\textrm{drag,top}})$, where $p_{\textrm{bottom}}$ is the bottom boundary of the simulation domain (200 bar). For the simulations presented here, $k_F=10^{-4}$ s$^{-1}$ and $p_{\textrm{drag,top}}=10$ bar. 

The key simulation parameters are summarized in Table \ref{tab:modelparameters}. HD~189733b has a radius of $1.13 \pm 0.01 R_J$ and a mass of $1.13 \pm 0.08 M_J$ \citep{StassunEtAl2017}, resulting in a gravity of 21.93 m s$^{-2}$. The orbital period of the planet is $2.21857567\pm0.00000015$ days. As it is expected that tidal forces have synchronized the rotation of the planet, we assumed this period to be the rotation of the planet around its axis. We further assume that the atmosphere behaves as ideal gas and assume values for the specific gas constant and the heat capacity appropriate for a hydrogen-dominated atmosphere (see Table \ref{tab:modelparameters}).

We initialize the simulation from a state of rest with a pressure-temperature profile based on a global-mean radiative equilibrium solution. We run the nominal simulations for 4,500 Earth days of simulation time and average the results over the last 100 days of simulation time.

\subsection{Radiative transfer}
\label{subsec:radtran}
We use the TWOSTR package \citep{KyllingEtAl1995} to solve the radiative transfer equations for a plane-parallel atmosphere in the two-stream approximation. This package is based on the general-purpose multistream discrete ordinate algorithm DISORT \citep{StamnesEtAl1988}. The opacities in the visible and the thermal band are set to constant values of $\kappa_{v}=6 \cdot 10^{-4} \sqrt{T_{\mathrm{irr}}/2000 \text{K}}$ m$^2$ kg$^{-1} =5.5 \cdot 10^{-4}$ m$^2$ kg$^{-1}$ and $\kappa_{\mathrm{th}}=10^{-3}$ m$^2$ kg$^{-1}$. These values were chosen based on Section 2.5 of \citet{Guillot2010RadiativeEquilibrium} who found that these values lead to a good match of their analytical solution for the temperature profile to the more detailed numerical models of \citet{FortneyEtAl2008}.
The assumption of gray opacities is a poor approximation for very low pressures ($p<$ 1 mbar), leading to significantly higher temperatures in these regions compared to models using more complex radiative transfer methods, such as SPARC \citep{ShowmanEtAl2009}. However, the circulation pattern remains qualitatively similar. The vertical velocities, crucial for the mixing of haze particles, show qualitatively similar patterns between the double-gray model and SPARC, though the peak velocities (restricted to localized regions near the terminators) can differ by up to a factor of two. The effect of higher temperatures at low pressures on the particle settling velocities is small.  For example, the settling velocities at temperatures of 800~K and 1,000~K differ only by 8\%.
Given that the focus of this paper is on the dynamical mixing rather than on the detailed thermal structure or emission spectra, the assumption of constant opacities is thus expected to be sufficiently accurate. Using gray opacities allows us to achieve longer simulation runtimes, and thus to explore the convergence of the haze distribution and the dependence on model parameters more thoroughly.

\subsection{Haze parametrization}
We utilize the passive tracer package of the MITgcm to simulate the production, advection and loss of photochemical hazes. For simplicity, we assume that the haze particles have a constant particle radius $a$ throughout the simulation domain, and vary the size of the particles as a free parameter. The haze mass mixing ratio $\chi$ follows the equation
\begin{equation}
\frac{D \chi}{Dt} = -g \frac{\partial (\rho \chi V_s)}{\partial p} + P + L,
\label{eq:tracereqn}
\end{equation}
where $D/Dt$ is the material derivative $\partial/\partial t + \bf{v}_H \cdot \nabla_H + \omega \partial/\partial p$, with $\bf{v}_H$ being the horizontal velocity, $\bf{\nabla}_H$ the horizontal gradient operator on a sphere in pressure coordinates and $\omega$ the vertical velocity in pressure coordinates. Furthermore, $\rho$ is the gas density, $V_s$ is the terminal velocity at which haze particles settle in the atmosphere in m s$^{-1}$ and $P$ and $L$ are the haze production and loss terms.
The terminal velocity is given by
\begin{equation}
V_s = \frac{2 \beta a^2 g (\rho_p-\rho)}{9 \eta},
\end{equation}
where $\beta$ is the Cunningham factor, $g$  the gravitational acceleration, $\rho_p$ the density of the particle, $\rho$ the gas density and $\eta$ the viscosity. We assume a value of $\rho_p=1,000$ kg m$^{-3}$, which is within the expected range of densities for particles with soot-like properties.

The Cunningham factor $\beta$ is a correction to the Stokes drag force in the regime where the mean free path of the gas is comparable to or larger than the particle size. The form of the Cunningham factor is determined experimentally. We adopt a form that has been recommended widely throughout the literature \citep[e.g.,][]{PruppacherKlett} and has also been used in previous exoplanet studies \citep{SpiegelEtAl2009,ParmentierEtAl2013}:

\begin{equation}
\beta =1 + Kn (1.256 + 0.4 e^{-1.1/Kn}),
\end{equation}
where $Kn=\lambda/a$ is the Knudsen number, defined as the ratio of the mean free path of the gas $\lambda$ to the particle size. 
Like \citet{ParmentierEtAl2013}, we use the parametrization given in \citet{Rosner2000} for viscosity:
\begin{equation}
\eta = \frac{5}{16}\frac{\sqrt{\pi m k_B T}}{\pi d^2} \frac{(k_B T/\epsilon)^{0.16}}{1.22},
\end{equation}
with the properties of molecular hydrogen (molecular diameter $d=2.827\cdot 10^{-10}$ m, molecular mass $m=3.34 \cdot 10^{-27}$ kg, depth of Lennard-Jones potential well $\epsilon=59.7 k_B$ K).

We assume that the production profile of photochemical hazes is a normal distribution in log-pressure space centered at 2 $\mu$bar. The width of the profile was chosen such that haze production is negligible in the two topmost layers. The production rate scales with the cosine of the zenith angle, $\theta$, of the incoming starlight. The production profile thus can be written as:
\begin{equation}
f (x,\theta) dx = \frac{F_0}{\sqrt{2 \pi} \sigma} \exp \left(- \frac{(x-x_0)^2}{2 \sigma^2} \right) \cos(\theta) dx,
\end{equation}
where $x=\log_{10} (p/1 \textrm{Pa})$, $x_0 = \log_{10}(0.2)$, $\sigma=0.25$ and $F_0=10^{-10}$ kg m$^{-2}$ s$^{-1}$ is the column-integrated mass production rate at the substellar point. The value of $F_0$ is consistent with the haze production mass fluxes used in \citet{LavvasKoskinen2017} (note that the fluxes reported in their paper are for a dayside-average, not for the substellar point).

In order to have a clearly defined bottom boundary condition, we assume that the hazes disappear at pressures higher than a threshold pressure, $p_{\mathrm{deep}}$. Such a deep sink can represent two processes: the condensation of cloud species on the haze particles, with the haze particles serving as condensation nuclei being lost from the population of `pure' photochemical hazes, and the possible thermal ablation of haze particles in deep, hot regions of the atmosphere. We represent both of these processes through the idealized sink term,
\begin{equation}
L=  
\begin{cases}
	0 & \text{for } p<p_{\mathrm{deep}}, \\
	-\chi/\tau_{loss} & \text{for } p>p_{\mathrm{deep}} ,
\end{cases}
\end{equation}
with the loss time scale $\tau_{\mathrm{loss}}=10^{3}$ s and $p_{\mathrm{deep}}=100$ mbar.

In early simulations, we found that numerical instabilities occurred in regions in which the settling velocity is large compared to the vertical velocity. This is the case at very low pressures and becomes worse for larger particle sizes. The instability can be suppressed by using an upstream difference scheme instead of a central difference scheme in the vertical derivative of the settling flux (first term on the right hand side of Eq. (\ref{eq:tracereqn})). However, upstream schemes have the disadvantage of introducing increased numerical diffusion. We therefore implemented a difference scheme that smoothly transitions from a central difference scheme, when the settling velocity is small compared to the vertical velocity, to an upstream scheme, when the settling velocity is large compared to the vertical velocity. This approach was inspired by comparable schemes that have been developed for solving the diffusion-advection equation \citep[e.g.,][]{FiadeiroVeronis1977,Wright1992}. The scheme used here was adapted from \citet{LavvasEtAl2010} with some modifications.
Details are presented in Appendix \ref{sec:appendixsettlingscheme}.

\section{Results}
\label{sec:results}
\begin{figure*}
\includegraphics[width=\textwidth]{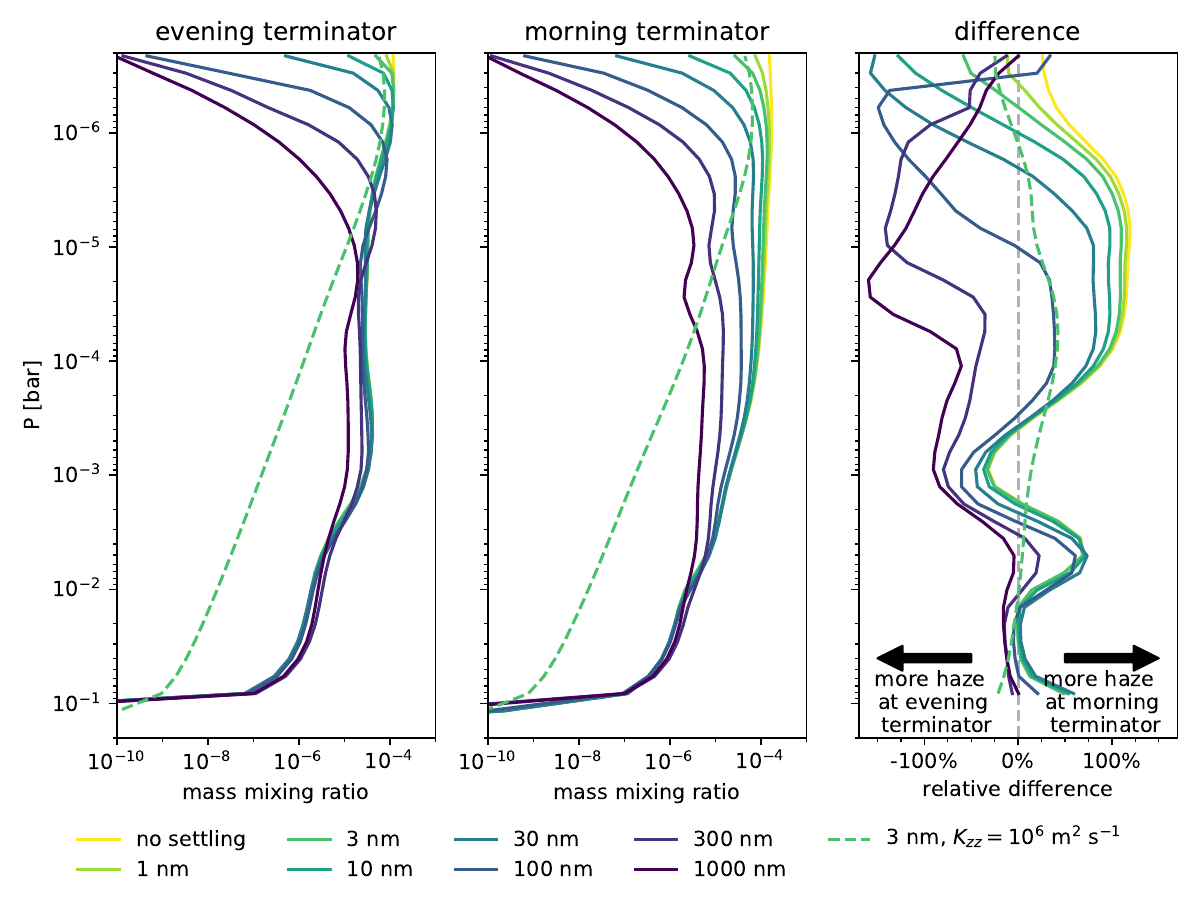}
\caption{Vertical profiles of the haze mass mixing ratio at the trailing (left panel) and leading limb (center panel). Profiles are time-averaged and averaged in latitude. The right panel shows the relative difference between the limbs, defined as $2 (\chi_{l}-\chi_{t})/(\chi_{l}+\chi_{t})$. } 
\label{fig:abundanceprofileplot_terminator}
\end{figure*}
\subsection{Overview of atmospheric circulation}
\label{subsec:circulationoverview}
The atmospheric circulation strongly shapes the distribution of photochemical hazes. Therefore, we start out with a brief description of the atmospheric circulation. At low pressures, the horizontal velocity field (arrows in Fig. \ref{fig:velocities}) is dominated by day-to-night flow. As pressure increases, the day-to-night component of the flow becomes weaker and the superrotation in the equatorial region becomes stronger. At $p>0.1$~mbar, the eastward equatorial jet becomes continuous.
On the nightside, at mid-to-high latitudes, a large vortex forms east of the antistellar point in each hemisphere. It extends past the morning terminator, slightly reaching into the dayside. The vortex is strongest at low pressures but extends through many orders of magnitude in pressure, up to $\approx100$~mbar.

The vertical velocity field (colorscale in Fig. \ref{fig:velocities}) is particularly important for mixing. The dayside is dominated by upwelling motion. The nightside contains both regions of upwelling and downwelling. There are two regions of particularly high vertical velocities extending over a large range of pressures: One is an eastward-pointing chevron-shaped feature near the morning terminator where strong downwelling follows strong upwelling (in the direction of the flow). A similar feature has also been observed in simulations using different GCMs, for example the model in \citet{FlowersEtAl2019} (Emily Rauscher, private communication). The feature was identified as hydraulic jump by \citet{ShowmanEtAl2009}.  A hydraulic jump is a transition from a state in which the flow velocity exceeds the propagation speed of gravity waves (supercritical flow) to a state in which the flow velocity is smaller than the wave propagation speed (subcritical flow) along the direction of flow. This transition usually is accompanied by strong vertical mixing.

The second region is a region of strong downwelling located west of the antistellar point, where flow coming from the dayside converges. This region of downwelling extends from the equator to midlatitudes. Near the equator, there is an adjacent region of upwelling that also extends over several orders of magnitudes in pressure.
These regions of particularly strong up- and downwelling consistent over several orders of magnitude in pressure were referred to as `chimneys' by \citet{ParmentierEtAl2013} and \citet{KomacekEtAl2019}.

\subsection{Small particle regime}
\label{subsec:smallparticles}
\begin{figure*}
\includegraphics[width=\textwidth]{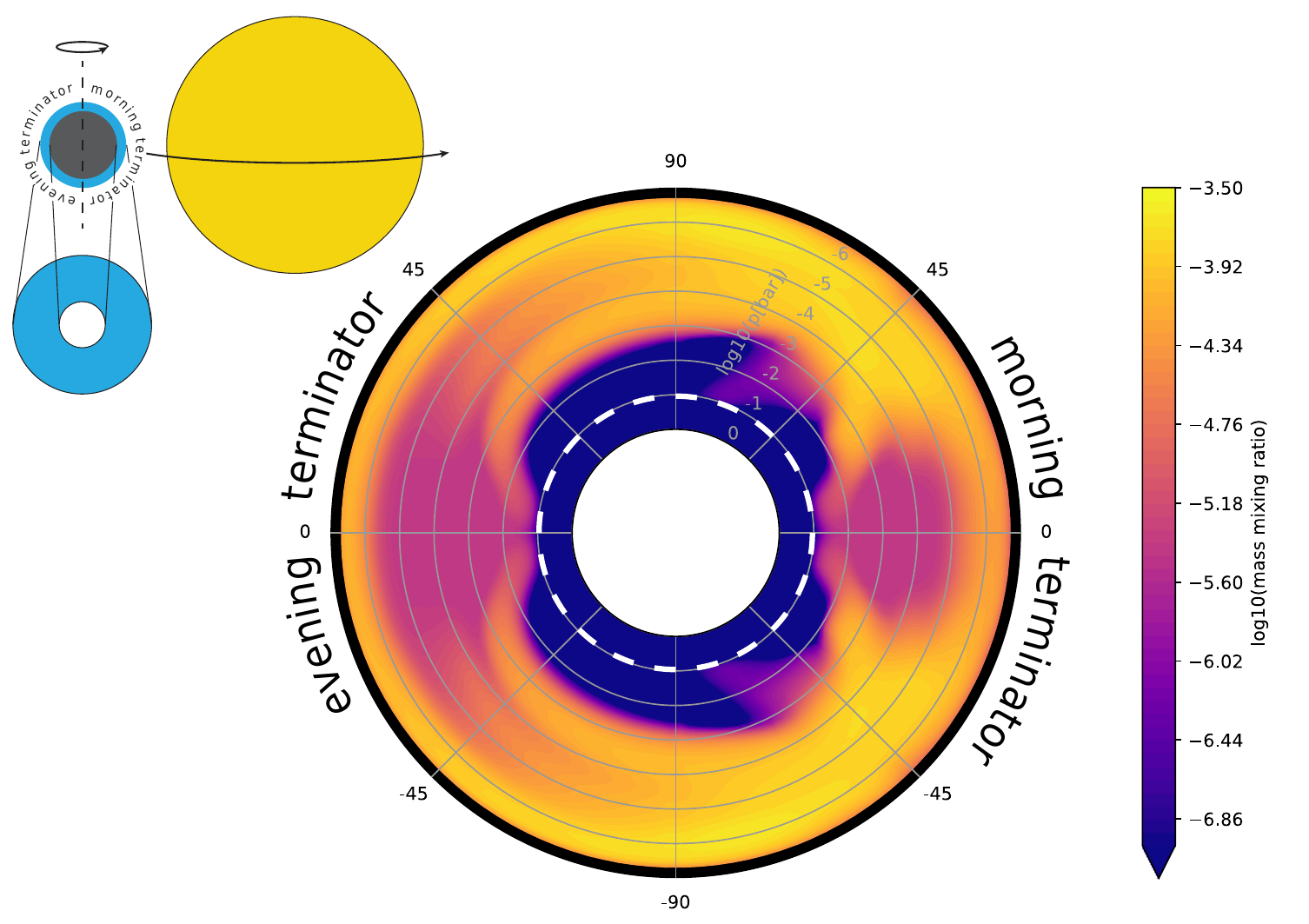}
\caption{Time average of the haze mass mixing ratio shown across a vertical slice of the atmosphere at the terminator for a particle size of 3 nm. The North pole is at the top and the morning terminator (leading limb) towards the right. The dashed white line indicates the pressure above which hazes are not allowed to exist.}
\label{fig:abundances_terminator_smallparticles}
\end{figure*}
For hazes with particle radii $\leq$10 nm, the settling time scale is longer than the advection time scale throughout the entire simulation domain. The three-dimensional distribution of the mass mixing ratio in this regime is thus determined by dynamical mixing and only weakly depends on the particle size. Indeed, our simulations show that the mass mixing ratio distribution is qualitatively similar for particle sizes $\lessapprox$10 nm (including the simulation in which particle settling is disabled). For particle sizes of 30 and 100 nm, the mass mixing ratio still looks qualitatively similar at pressures larger than $\approx$10 $\mu$bar. This can be seen in Fig. \ref{fig:abundanceprofileplot_terminator}, which shows the mass mixing ratio profiles averaged across the terminator for different particle sizes. The yellow and green colors correspond to the small particle regime in this figure. For the remainder of this section, we show results for a particle radius of 3 nm in our figures. This radius is close to the mean particle size at low pressures predicted by microphysics models \citep{LavvasKoskinen2017}.

Because our main interest lies in the haze distribution at the terminator, as seen in transit spectroscopy, we start with a description of that. Figure \ref{fig:abundances_terminator_smallparticles} shows the haze mass mixing ratio along a cross-section of the terminator for a particle size of 3 nm. Contrary to the predictions by \citet{KemptonEtAl2017}, in general, there is a higher mass mixing ratio at the morning terminator (leading limb in transit, west of the substellar point) than at the evening terminator (trailing limb in transit, east of the substellar point). This difference is particularly pronounced for pressures lower than a few mbars. For these low pressures, the mixing ratio is highest near the poles and at midlatitudes at the morning terminator. At the evening terminator, mixing ratios are uniformly low throughout the equatorial region and midlatitudes. At pressures larger than 1 mbar, there are enhanced mixing ratios at low latitudes, separated by an equatorial band that is depleted in hazes. There are also enhanced mixing ratios near the poles. At midlatitudes, there is a strong depletion of hazes for these pressures. To better understand this picture, we now move on to describe the three-dimensional distribution of hazes.

\begin{figure}
\includegraphics[width=\columnwidth]{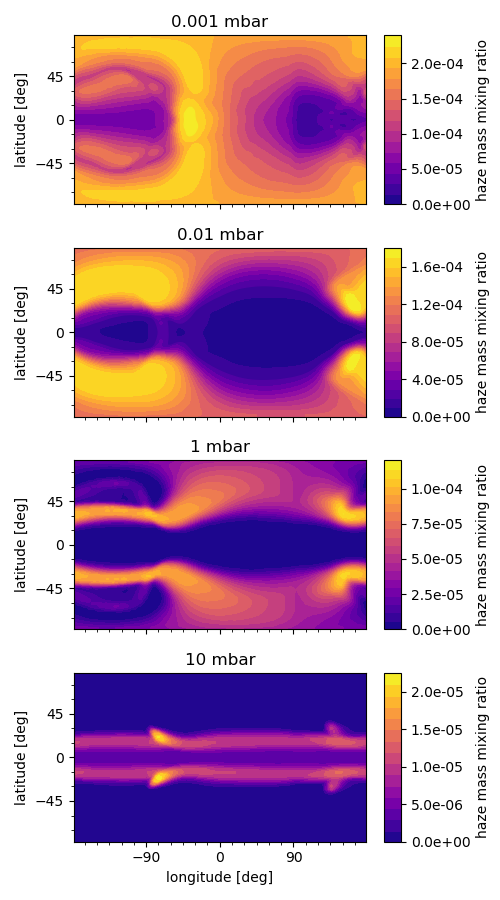}
\caption{Time-averaged haze mass mixing ratio at several isobars for a particle size of 3 nm. As in Fig. \ref{fig:velocities}, the substellar point is at the center of the panel.}
\label{fig:smallparticles_isobars}
\end{figure}

At altitudes near and above the haze production peak, the highest mixing ratio is west of the substellar point. Upwelling on the dayside lifts haze particles above the  peak production level. The day-to-night flow then carries particles polewards and towards the nightside. In the equatorial region, there is westward flow on large fractions of the dayside (west of $\approx45^\circ$ longitude) at these low pressures, causing the mixing ratio to peak west of the substellar point. This can be seen in the top panel of Fig. \ref{fig:smallparticles_isobars}.

Haze particles then are mixed downwards on the nightside, where they get trapped in the midlatitude vortices. Between 3 $\mu$bar and $\approx0.1$ mbar,  there are clearly enhanced mixing ratios in these vortices which extend over large parts of the nightside. This explains the higher abundances at the morning terminator in this pressure range compared to the evening terminator: the regions of enhanced abundances associated with the two midlatitude vortices extend across the morning terminator, reaching slightly into the dayside. At the evening terminator, in contrast, there is upwelling, mixing up air depleted of haze, as well as flow from the dayside which is also dominated by upwelling.

\begin{figure}
\includegraphics[width=\columnwidth]{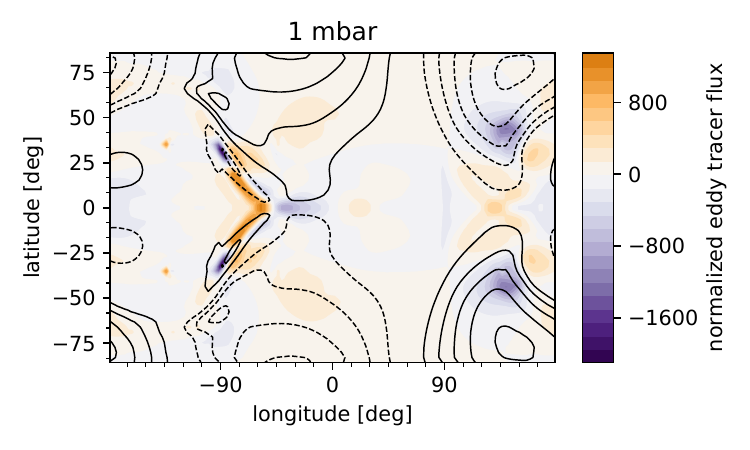}
\caption{Vertical and meridional transport at the 1 mbar level: The color scale represents the normalized eddy tracer flux $w \chi' /\vert\overline{w \chi'}\vert$, where  $\chi'=\chi-\overline{\chi})$ and the bar represents the horizontal average of a quantity. Positive values of the normalized eddy tracer flux indicate upward transport of the tracer. The contours show the meridional velocity in intervals of 500 m/s, with dashed lines indicating negative (southward) velocities.}
\label{fig:eddytracerflux}
\end{figure}

At pressures between 0.1 mbar and 1 mbar, the horizontal distribution of hazes gradually transitions to a more longitudinally symmetric pattern with increasing pressure. This transition is likely caused by the equatorial jet becoming more dominant with increasing pressure, while the day-to-night-flow component becomes much weaker. On each side of the equator, a band with enhanced haze abundance forms. The band widens  and moves to slightly higher latitudes on the dayside and narrows and moves closer to the equator on the nightside. There are two longitudes at which the abundance within this band peaks: near the morning terminator, caused by the hydraulic jump described earlier, and near the antistellar point, caused by converging flow leading to a narrowing of the band and downwelling. With increasing pressure, the band moves closer to the equator. To better understand why the band is moving closer to the equator with increasing pressure, we examined the vertical eddy tracer flux $w(\chi-\overline{\chi})$ (Fig. \ref{fig:eddytracerflux}), where $w$ is the vertical velocity and $\overline{\chi}$ denotes the horizontally averaged mass mixing ratio.  The vertical eddy tracer flux is a measure for the local strength and direction of vertical transport. We find that in the pressure regions in which the bands form, there is a highly localized peak of the downward eddy tracer flux at the location where the morning terminator hydraulic jump intersects with the band of high tracer abundance. The morning terminator hydraulic jump feature also exhibits equatorward meriodional velocities. Therefore, at this location, air enriched in hazes is simultaneously transported downwards and equatorwards. Similarly, at the downwelling region slightly west of the antistellar point, equatorward meridional velocities prevail. We suggest that this causes the band of high haze abundance to move equatorward with increasing pressure. 

We further note that the equatorial region is almost uniformly depleted in hazes at all longitudes for 3 $\mu$bar$<p<10$ mbar. This cannot be explained by a simple correlation with upwelling and downwelling, as there are both regions of upwelling and downwelling at the equator.  We point out that this equatorial depletion is distinct from the equatorial depletion observed in simulations of condensate clouds in hot Jupiter and mini-Neptune atmospheres \citep{ParmentierEtAl2013,CharnayEtAl2015a,LinesEtAl2018}, which happens at much higher pressures. The latter has been suggested to be caused by downwelling at the equator in the zonal-mean circulation \citep{CharnayEtAl2015a}. However, for photochemical hazes produced high in the atmosphere, downwelling is expected to cause enhanced abundances. Therefore, the depletion observed in our simulations cannot be explained by the same mechanism.  At the same time, because these two different kinds of depletion happen at very different pressures, our result does not contradict the previous studies.

\begin{figure*}
\includegraphics[width=\textwidth]{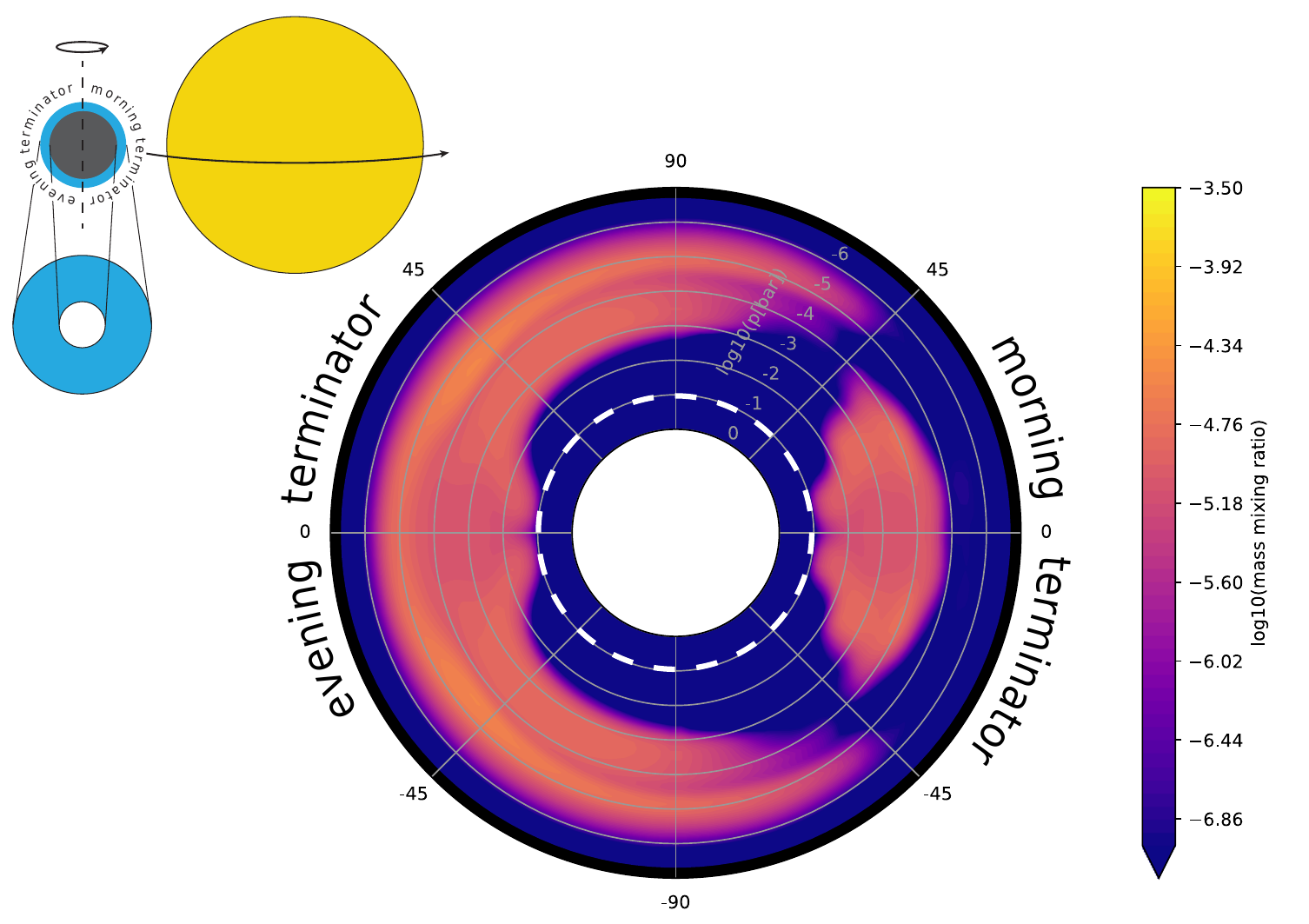}
\caption{Time average of the haze mass mixing ratio shown across a vertical slice of the atmosphere at the terminator for a particle size of 1 $\mu$m. The North pole is at the top and the morning terminator (leading limb) towards the right. The dashed white line indicates the pressure above which hazes are not allowed to exist.}
\label{fig:abundances_terminator_largeparticles}
\end{figure*}

To summarize, we find that near and above the peak haze production region ($p\lessapprox3 \mu$bar), hazes are transported towards the nightside by the direct day-to-night flow. Downwelling on the nightside leads to enhanced haze mixing ratios on the nightside for ($3 \mu$bar$\lessapprox p\lessapprox0.1-1$ mbar), where they accumulate in the two large cyclonic vortices at midlatitudes. Because the vortices extend across the morning terminator, there are enhanced haze mixing ratios at the morning terminator for that pressure range. Upwelling throughout large parts of the dayside and near the evening terminator correlate with low abundances at the evening terminator. Deeper in the atmosphere ($p\gtrapprox0.1-1$ mbar), the mass mixing ration distribution becomes more longitudinally symmetric.

\subsection{Large particle regime}
\begin{figure}
\includegraphics[width=\columnwidth]{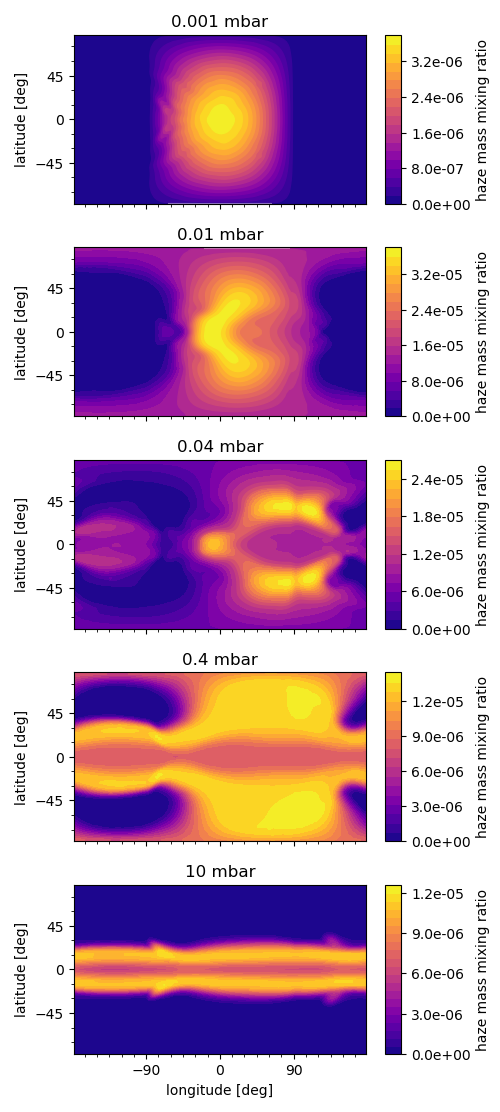}
\caption{Time-averaged haze mass mixing ratio at several isobars for a particle size of 1 $\mu$m. As in Fig. \ref{fig:velocities}, the substellar point is at the center of the panel.}
\label{fig:largeparticles_isobars}
\end{figure}

In the large particle regime ($a > 30$ nm), the 3D mass mixing ratio pattern depends much more strongly on the particle size. Key factors determining the resulting pattern are the ratio of the settling time scale $\tau_s$ to the advection time scale $\tau_{\mathrm{adv}}$ and the wind pattern at the level at which the settling timescale becomes comparable to the advection time scale. Therefore, it is harder to describe the general behavior of particles in that regime instead of describing each particle size separately. However, some patterns can be identified.

We again start with the haze distribution at the terminator (shown for a particle size of 1~$\mu$m in Fig. \ref{fig:abundances_terminator_largeparticles}).
In the large particle regime, there are overall higher mass mixing ratios at the evening terminator than at the morning terminator, as suggested by \citet{KemptonEtAl2017}. At very low pressures($p\lessapprox9 \cdot10^{-7}$ bar for 300~nm, $p\lessapprox3 \cdot 10^{-6}$~bar for 1~$\mu$m), there are hardly any hazes present at any part of the terminator. Below that pressure region, hazes are present at all latitudes at the evening terminator, with the mixing ratio peaking at mid- to high latitudes. At the morning terminator, in contrast, hazes are only present at high latitudes. Then there is a pressure region, in which hazes are present throughout the entire terminator (evening and morning side) except for a region at midlatitudes at the morning terminator ($10^{-5}\lessapprox p\lessapprox 10^{-3}$~bar for 300~nm, $5 \cdot 10^{-5}\lessapprox p \lessapprox 10^{-3}$~bar for 1~$\mu$m). Finally, deeper in the atmosphere ($p\gtrapprox 10^{-3}$\~bar), hazes are concentrated in two bands at low latitudes at both the evening and morning terminator, with slightly lower abundances in between those bands at the equator and no hazes present at higher latitudes. This general picture applies to all  particle sizes within the large particle regime, though the pressure regions described move to lower pressures for smaller particles and to higher pressures for larger particles due to the different settling timescales. As in the small particle regime, the terminator differences can be understood by looking at the three-dimensional haze distribution.

At very low pressures, where $\tau_{s}/\tau_{\mathrm{adv}}\ll 1$, hazes are concentrated on the dayside with a pattern closely resembling the production function, slightly modulated by the vertical velocity pattern (top panel of Fig. \ref{fig:largeparticles_isobars}). Only where both timescales become comparable ($0.1 \lessapprox\tau_s/\tau_{\mathrm{adv}} \lessapprox 1$), the winds significantly change the haze pattern. A westward pointing chevron emerges at the dayside (second panel from top). The pattern is caused by increasingly eastward winds at midlatitudes on the dayside below the haze production region (opposed to the polewards and westwards flow near the peak production region shaping the distribution of hazes in the small particle regime--compare the left three panels of Fig. \ref{fig:velocities}). A few layers deeper in the atmosphere, hazes are mainly found on the hemisphere east of the substellar point, with two symmetric arcs with a higher concentration of hazes that connect the substellar point to the regions of strong downwelling west of the antistellar point (third panel). There also is a somewhat lower but significant haze mass mixing ratio at high latitudes all around the globe, transported there by the day-to-night flow. Moving on to higher pressures, hazes become increasingly concentrated at these nightside downwelling spots. There also is an increasing amount of hazes at low latitudes on the nightside and the morning terminator, comparable to that near the poles. The nightside midlatitude vortices remain severely depleted of hazes. Once $\tau_s/\tau_{\mathrm{adv}} $ reaches values larger than unity, there is a pressure region in which hazes are distributed relatively uniformly around the globe except at the nightside vortices (fourth panel). At pressures larger than 1 mbar, where the equatorial jet dominates the circulation and $\tau_s/\tau_{\mathrm{adv}} \gg 1$, a banded pattern comparable to the one seen in the small particle regime appears (bottom panel).

\section{Comparison to 1D Models: Deriving an effective eddy diffusion coefficient}
\label{sec:eddydiffusioncoefficient}
\begin{figure}
\includegraphics[width=\columnwidth]{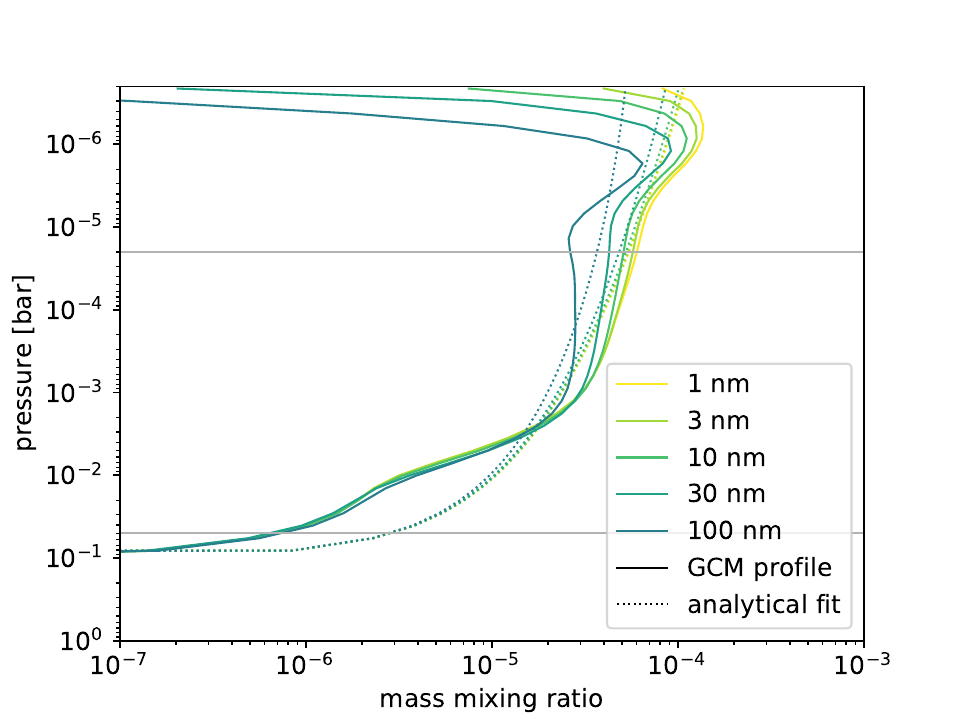}
\caption{Comparison of the analytical solution to the 1D haze transport equation using the K$_{zz}$ parametrization in Eq. (\ref{eq:Kzz_smallparticles})  (dotted lines) to the horizontally-averaged mass mixing ratio profile from the 3D simulation (solid lines). The gray horizontal lines indicate the region over which the fit used to derive Eq. (\ref{eq:Kzz_smallparticles}) was performed.}
\label{fig:analyticalsolution_smallparticles}
\end{figure}

\begin{figure}
\includegraphics[width=\columnwidth]{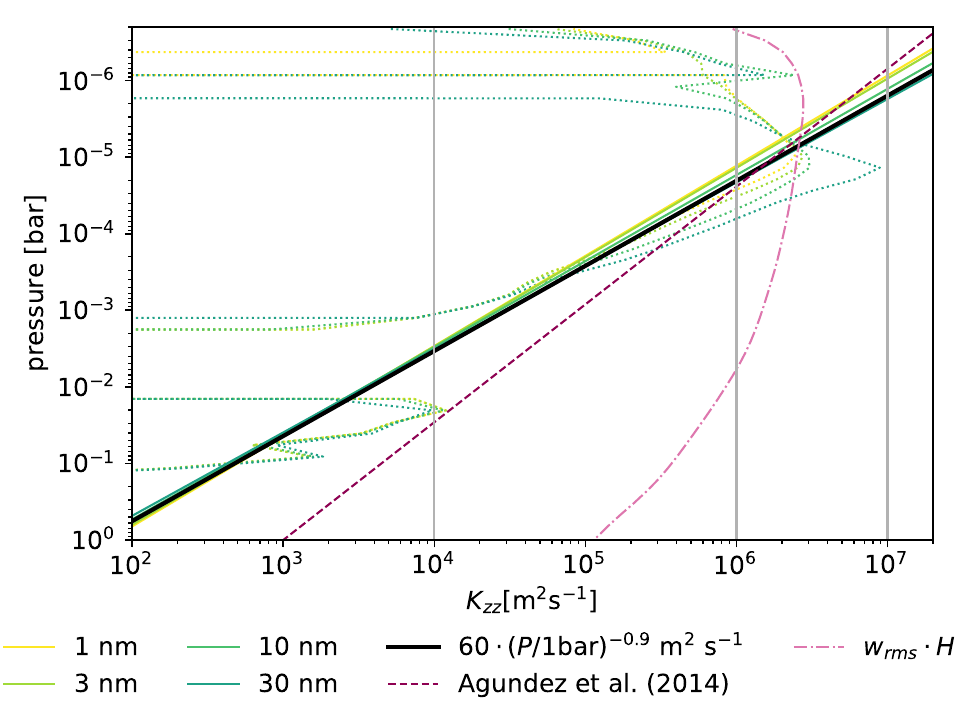}
\caption{Comparison of effective eddy diffusion coefficients derived using different methods. The dotted lines show eddy diffusion coefficients derived using Eq. (\ref{eq:kzz_eddytracerflux}). Solid lines represent the fit to the analytical solution, with yellow and green lines representing different particle sizes and the black line the suggested `average' parametrization for all particle sizes (Eq. \ref{eq:Kzz_smallparticles}). The magenta dashed line shows the equivalent power-law derived by fitting to the analytical solution for cloud-like tracers from \citet{AgundezEtAl2014}. The pink dot-dashed line indicates the common estimate for K$_{zz}$ using the root-mean-square vertical velocity and pressure scale height. This profile is similar (but not identical) to the profile used in \citet{MosesEtAl2011}, which was derived from the output of the GCM simulations of \citep{ShowmanEtAl2009} using the same method. The three $K_{zz}$ values used in the simulations with enhanced sub-grid-scale mixing shown in Fig. \ref{fig:enhancedkzz_terminator} are indicated by the gray vertical lines.  }
\label{fig:kzzcomparison}
\end{figure}

After describing the 3D distribution from our simulations, we now turn to the strength of vertical mixing and how it compares to assumptions used in 1D models. In 1D models, vertical mixing usually is parametrized by an eddy diffusion coefficient $K_{zz}$. In this section, we derive an effective eddy diffusion coefficient that describes the haze distribution observed in our simulation.

There are multiple methods for deriving effective eddy diffusion coeffcients from a GCM. For cases in which an analytical solution to the one-dimensional transport equation exists, a good method is to derive $K_{zz}$ by fitting the analytical solution to the horizontally-averaged mass mixing ratio profile. We aim therefore to formulate a simple problem for which analytical solutions can be found for the stationary case and that provides a 1D-analog to our numerical simulations.

In the absence of sources and sinks, the 1D equivalent to Eq. (\ref{eq:tracereqn}) is
\begin{equation}
    \frac{\partial \chi}{\partial t} - g^2 \frac{\partial}{\partial P} \left( \rho^2 K_{zz} \frac{\partial \chi}{\partial P} \right) = - g \frac{\partial(\rho \chi V_s)}{\partial P}.
\end{equation}
 To mimic haze production at a high altitude, we impose a fixed downward mass flux $F_0$ at the low-pressure boundary of the domain $P_1$. At the high-pressure boundary $P_0$, we force the tracer mixing ratio to be zero.

Assuming that $K_{zz}$ can be described by a power law, $K_{zz}=K_{zz,r} (P/P_r)^{-\alpha}$, and that the settling velocity is inversely proportional to pressure (a valid approximation for $Kn\gg 1$), $V_s=V_r (P/P_r)^{-1}$, the analytical solution then is given by
\begin{equation}
    \chi = \frac{F_0 g H}{V_r P_r} \left\{ 1 - \exp \left[\frac{V_r H P_r^{1-\alpha}}{K_{zz,r}} \frac{1}{\alpha-1} \left(P^{\alpha-1}-P_0^{\alpha-1}\right) \right] \right\},
\end{equation}
where $P_r$ is a reference pressure.
We fitted the analytical solution to the horizontally-averaged tracer mass mixing ratio profile for each particle size separately using the curve\_fit function of the scipy optimize module. The fit was performed over a pressure range from 50 mbar to 20 $\mu$bar (below the haze production region). For particle sizes $\leq 100$ nm, we obtain very similar results, with $\alpha$ ranging from 0.85 to 0.95 m$^2$s$^{-1}$  and $K_{zz,r}$ ranging from 45 to 70 for $P_r=1$ bar. We therefore suggest the following parametrization for small photochemical hazes in 1D models:
\begin{equation}
    K_{zz} = 60 \cdot \left(\frac{P}{1 \textrm{bar}}\right)^{-0.9} \mathrm{m}^2 \mathrm{s}^{-1}
    \label{eq:Kzz_smallparticles}
\end{equation}
A comparison of mass mixing ratio profiles using the analytical solution with this parametrization to the GCM results  is shown in Fig. \ref{fig:analyticalsolution_smallparticles}. 
We note that strictly speaking, the parametrization is only valid for the pressure range over which the fit has been performed.

For the largest two particle sizes in our simulations, 300 nm and 1 $\mu$m, the fit returned very different $K_{zz}$ profiles: For 300 nm, the best fit is $\alpha=0.33$ and $K_r=700$ m$^2$s$^{-1}$  while for 1 $\mu$m,  $\alpha=0.06$ and $K_r=2,000$ m$^2$s$^{-1}$ provide the best fit (though any value below 0.1 results in a similarly good fit).
This result for large particle sizes should be viewed with caution, however: For these particle sizes, gravitational settling dominates over vertical mixing/diffusion over a large pressure range (p$\lessapprox 1$ mbar for 300 nm, p$\lessapprox 5$ mbar for 1 $\mu$m) for typical values of $K_{zz}$. A different eddy diffusion coefficient thus only changes the analytical solution in the higher-pressure part of the atmosphere. Effectively, this means that the fit only uses a region spanning one order of magnitude in pressure to constrain $K_{zz}$ rather than over three orders of magnitude. Moreover, the region in which the analytical solution is sensitive to $K_{zz}$ for the largest two particle sizes is also the region in which the analytical solution for small particles shows the largest discrepancies from the numerical solution (see Fig. \ref{fig:analyticalsolution_smallparticles}). It seems more likely that the result for large particle sizes is therefore driven by the small size of the region the fit is sensitive to rather than implying that a different eddy diffusivity should be used to describe the mixing of large haze particles.

A second method for deriving $K_{zz}$ from the tracer distribution is based on the ratio of the vertical eddy tracer flux and the tracer gradient,
\begin{equation}
    K_{zz} = - \frac{\overline{w \chi'}}{\frac{\partial \overline{\chi}}{\partial z}},
    \label{eq:kzz_eddytracerflux}
\end{equation}
where the bar denotes a horizontal average of a quantity and the prime denotes the deviation of the quantity from its horizontal average.
For the small-particle regime, the results of both methods are consistent with each other, as can be seen in Fig. \ref{fig:kzzcomparison}. We note that $K_{zz}$ profiles derived using the second method tend to be less smooth because more local variations in the mass mixing ratio profile can cause spikes and dips.

For comparison, the profile used by \citet{AgundezEtAl2014} for HD~189733b is $K_{zz}=10^3 \cdot (P/1 \textrm{bar})^{-0.65}$ m$^2$ s$^{-1}$ (magenta dashed line). This profile was derived by fitting an analytical solution of the mass mixing ratio profile to the horizontally-averaged mass mixing ratio profile from 3D tracer simulations of condensate clouds. It is the equivalent to Eq. (22) in \citet{ParmentierEtAl2013} for an atmosphere with no thermal inversion and the planetary parameters of HD~189733b. The key difference to our simulation is that because their work focused on condensate clouds, no particle source at low pressures and a sink deep in the atmosphere were included. Instead, simulations were started with a uniform tracer abundance throughout the simulation and continued to run until a quasi-steady state had been achieved. Further, in their simulations, the tracer was subjected to gravitational settling on the night side only because the cloud species was assumed to be gaseous on the day side. Particle sizes in their simulations ranged from 100 nm to 10 $\mu$m. Unlike in this work, they were able to fit large particle sizes well because the solution for the situation without a low-pressure source term is more sensitive to the strength of vertical mixing.

We also show the commonly used estimate $w_{\mathrm{rms}} \cdot H$ (pink dot-dashed line). \citet{ParmentierEtAl2013} found that this prescription overestimates $K_{zz}$ by about two orders of magnitude. Our results also find that this prescription leads to much stronger vertical mixing than in the GCM. However, we also find a stronger pressure dependence of $K_{zz}$ in our simulations. Therefore, our derived $K_{zz}$ and $w_{\mathrm{rms}} \cdot H$ are within an order of magnitude at very low pressures but differ by three orders of magnitude at 100 mbar.

\section{Transmission spectra}
\label{sec:spectra}
In this section, we aim to explore the effect of the differences in haze abundance between the morning and evening terminator on the transmission spectra, which potentially could be measured through ingress and egress transit measurements by future instruments. We use a one-dimensional code \citep[as in][]{LavvasKoskinen2017} to calculate transmission spectra. As input, separate number density and temperature profiles averaged across the morning and evening terminator were used. For normalizing the spectra, the average between the transit radii calculated for morning and evening terminator is used (representing the observed transmission spectrum during mid-transit).
The transit spectra are normalized such that the transit radius of each model spectrum in the Spitzer 3.6 $\mu$m band is consistent with the observed radius in that band. This normalization is necessary because of the degeneracy between the assumed planet radius in the calculation and the associated reference pressure. For more details on the normalization procedure we refer to Section 3.3 in \citet{LavvasKoskinen2017}, noting that they use the 8 $\mu$m band for normalization instead of the 3.6 $\mu$m band.

\begin{figure}
\includegraphics[width=\columnwidth]{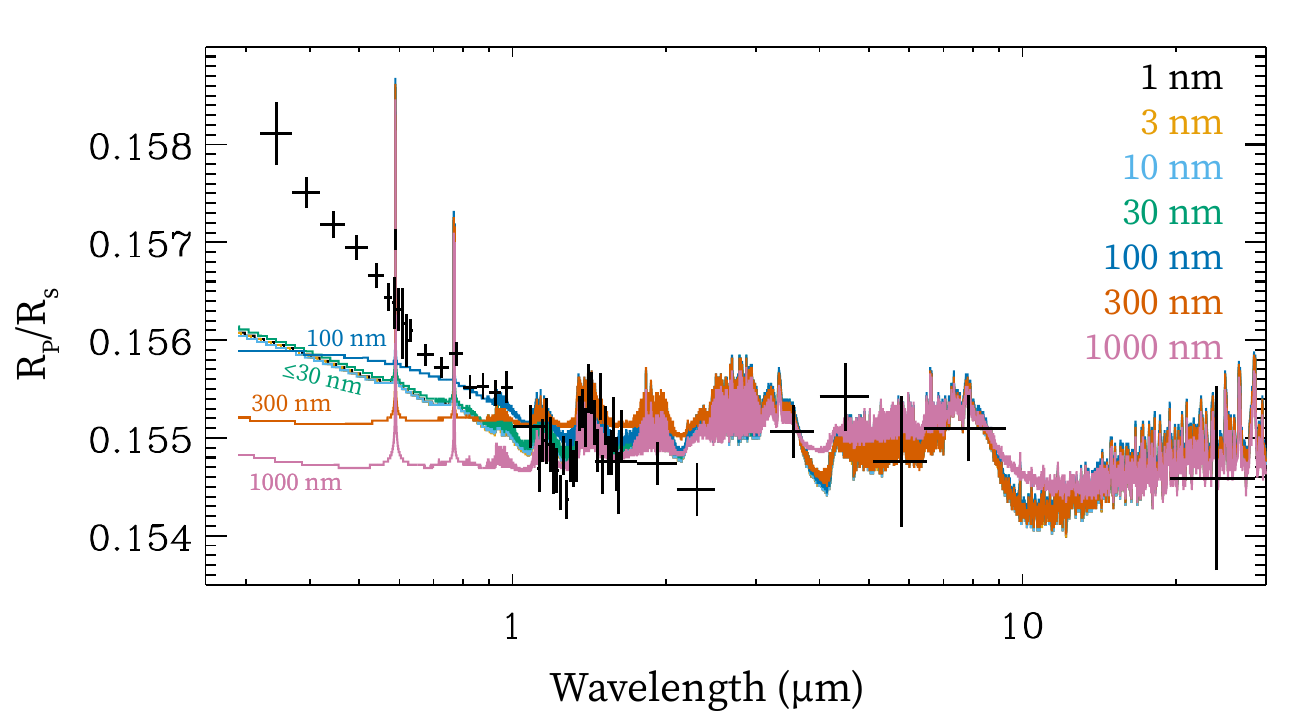}
\caption{Model transmission spectra for different particle sizes (colored lines, see text for details). Note that the spectra for all particle sizes $\leq$ 30 nm lie on top of each other. The black crosses represent observational data from \citet{SingEtAl2016} ($\lambda< 5\mu$m) and \citet{PontEtAl2013} ($\lambda>5 \mu$m).}
\label{fig:spectra_allsizes}
\end{figure}

There is a large uncertainty in the magnitude of the haze production rate. Because we use passive tracers which have no feedback on the atmospheric circulation and because all terms in Eq. (\ref{eq:tracereqn}) except for the production term are linear in $\chi$, it is possible to change the haze production rate without rerunning the simulation by simply scaling the mass mixing ratio and number density from the simulation output by the same constant factor as the haze production rate. We therefore choose to constrain the haze formation rate based on existing observations before looking at terminator differences. This is achieved by matching the model spectrum to observations in the spectral region of the \textit{HST WFC3} G141 grism (1.1-1.7 $\mu$m). 
Figure \ref{fig:spectra_allsizes} shows simulated transmission spectra for all particle sizes considered.
For all particle sizes except 1 $\mu$m, number densities from the nominal simulation were multiplied by a factor of 0.025 (corresponding to a column-integrated haze production rate of $2.5 \cdot 10^{-12}$ kg m$^{-2}$ s$^{-1}$ at the substellar point). This haze formation rate provides a reasonable match to the water feature within the \textit{WFC3} G141 spectral range for these particle sizes and will be used below to examine terminator differences. For particles with a radius of 1 $\mu$m, number densities were multiplied by a factor of 0.1 (corresponding to $10^{-11}$ kg m$^{-2}$ s$^{-1}$). We note that we are not able to match the slope at short wavelengths. This will be discussed further below. First, however, we explore the differences between the morning and evening terminator, focusing on the small particle regime.

\begin{figure}
\includegraphics[width=\columnwidth]{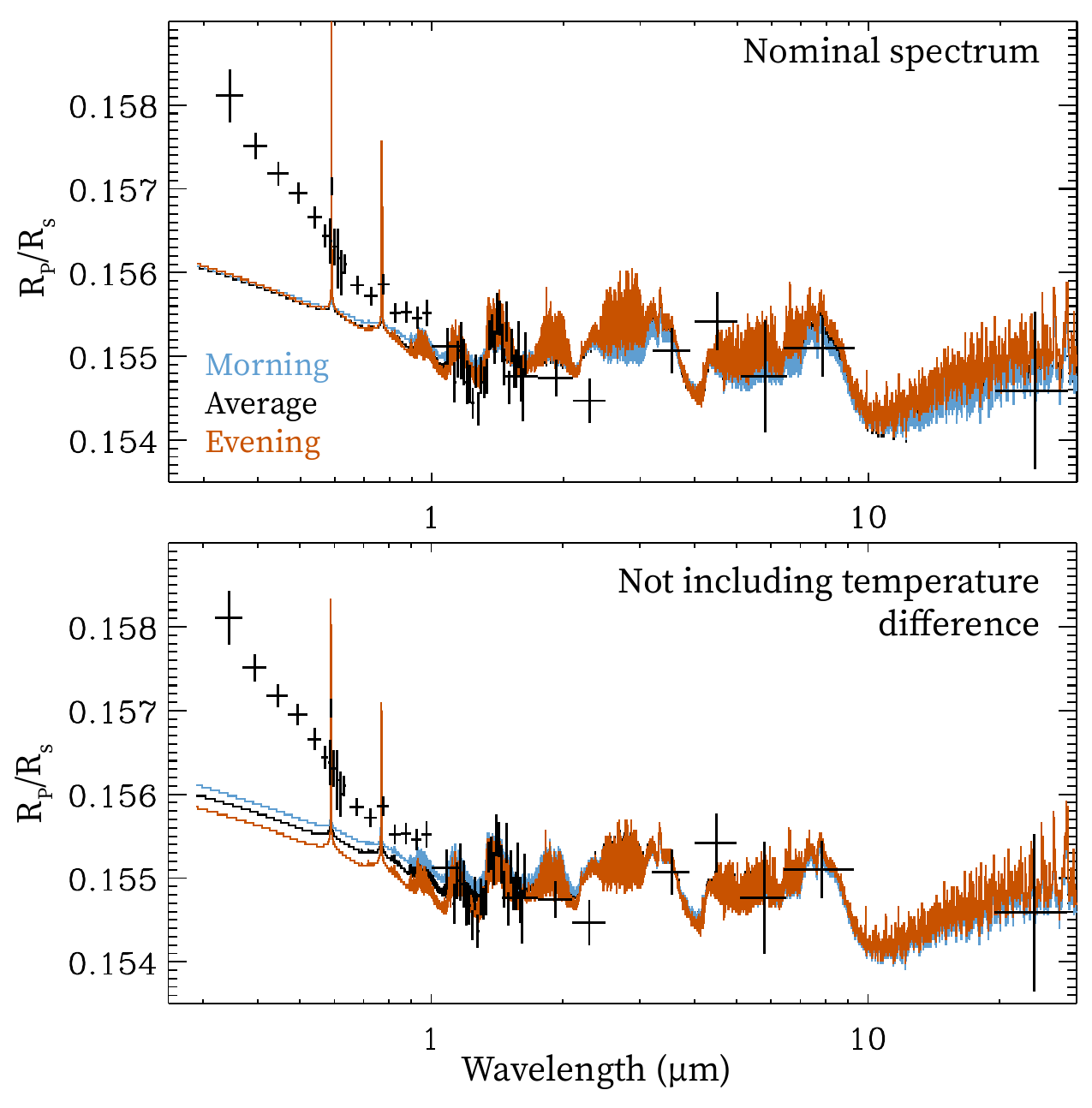}
\caption{Model transmission spectra for a particle radius of 3 nm, calculated for the morning (light blue) and evening (dark orange) terminator separately, as well as for the entire planet (black). In the top panel, consistent temperature profiles from the GCM simulation were used, with a colder morning terminator and a hotter evening terminator. In the bottom panel, an identical temperature profile was used for morning and evening terminator. The black crosses again represent observational data.}
\label{fig:spectra_3nm}
\end{figure}

The resulting differences in the transit radius between evening and morning terminator are very small. This can be seen in the top panel of Fig. \ref{fig:spectra_3nm} for the 3 nm case. The reason for this small difference appears to be that the two competing effects of differences in haze abundance and temperature offset each other. At the morning terminator, more hazes are present, increasing the transit radius. However, temperatures are also lower by $\approx 200-250$ K at the morning terminator in our simulation, resulting in a smaller scale height and thus a smaller transit radius. Both effects appear to roughly outweigh each other. 
To isolate the effect of the haze abundance, we recomputed the transit spectra using an identical temperature profile for evening and morning terminator (bottom panel of Fig. \ref{fig:spectra_3nm}). In this case, the transit radius at the morning terminator is significantly larger than at the evening terminator. This confirms that terminator differences in the nominal spectrum are so small because temperature difference and difference in haze abundance largely cancel each other.

We also explored the impact of using a particle size distribution instead of a single particle size. We used a particle size distribution from the 1D microphysics model of \citet{LavvasKoskinen2017} and interpolated the number densities from the GCM between the coarser particle size grid used in the GCM. The resulting transit spectrum (not shown) has a slightly shallower slope at short wavelengths. The differences between morning and evening terminator are similarly small as for single particle sizes within the small particle regime.

\section{Models with enhanced sub-grid-scale mixing}
\label{sec:extrakzz}
As we noted above, the slope in our model spectra is much shallower than the observed slope. While star spots could account for part of the slope \citep{McCulloughEtAl2014}, it seems unlikely that they can explain all of it. In this section, we therefore explore another scenario that could explain the steep slope. In one-dimensional models, a high eddy diffusion coefficient can lead to a stronger spectral slope \citep{LavvasKoskinen2017}.
\citet{OhnoKawashima2020} show that steep spectral slopes are possible with high eddy diffusion coefficients and intermediate haze production rates. In contrast, as discussed in Section \ref{sec:eddydiffusioncoefficient}, vertical mixing in our 3D simulations has a much stronger pressure-dependence, resulting in significantly weaker mixing at intermediate and higher pressures. Our nominal simulations presented so far, however, only capture mixing through the large-scale circulation and cannot resolve mixing by turbulent motion on a scale comparable to or smaller than the grid size. Such turbulent motion could be generated  by a variety of processes, including shear instability, atmospheric waves, disturbances at the convective-radiative boundary propagating upward, magnetic field effects or tidal waves. While some of these mixing processes could in theory be resolved by using a much higher (not feasible) model resolution, many other processes would further require us to add additional physics to our GCM or switch to a different code. For example, turbulence generated at the radiative-convective boundary would require additional forcing to be added near the bottom boundary of the model. Magnetic field effects would require a magneto-hydrodynamic code.
There are little constraints on the magnitude of such sub-grid-scale turbulent mixing and it is possible that its role is comparable to or larger than that of mixing by the large-scale circulation. To explore this possibility, we ran additional simulations in which we added additional mixing parametrized through an eddy diffusion coefficient acting on the passive tracers. These additional simulations were started from a converged simulation output from the nominal simulations and integrated for 500 days. The simulations reached a quasi-steady state within 200 to 300 days. As with the nominal simulations, the results presented here are based on the time-average of the last 100 days of integration. We ran simulations for 3 nm-sized particles for four different constant values of $K_{zz}$ ($10^3$ m$^2$s$^{-1}$,$10^4$ m$^2$s$^{-1}$,$10^5$ m$^2$s$^{-1}$,$10^6$ m$^2$s$^{-1}$, $10^7$ m$^2$s$^{-1}$) as well as for the pressure-dependent $K_{zz}$ profile of \citet{MosesEtAl2011} and the \citet{MosesEtAl2011} profile scaled by a factor of 0.01. In addition, we ran simulations with a constant $K_{zz}$ of $10^6$ m$^2$s$^{-1}$ for all particle sizes.

\subsection{3D distribution}
\begin{figure*}
\includegraphics[width=\textwidth]{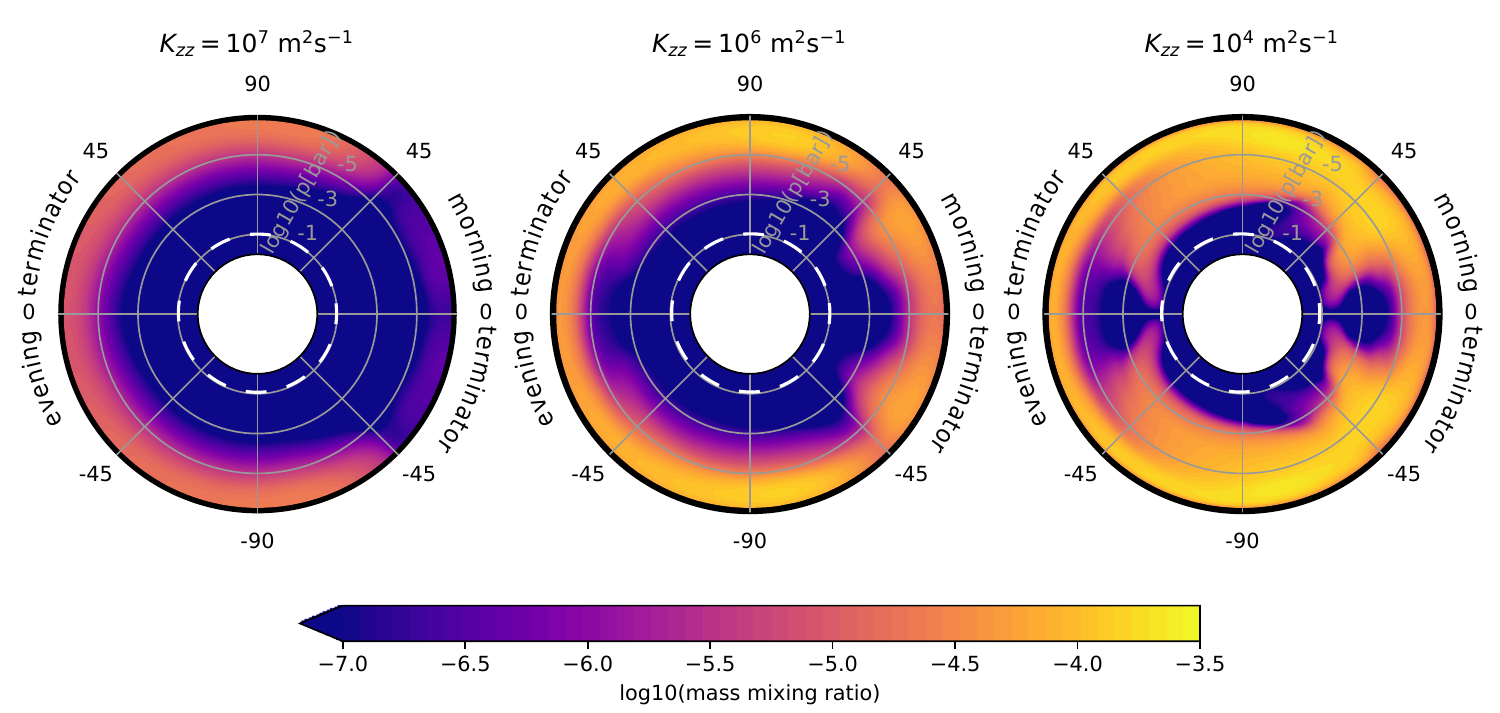}
\caption{Time average of the haze mass mixing ratio shown across a vertical slice of the atmosphere at the terminator for a particle size of 3 nm and different values of the eddy diffusion coefficient representing sub-grid-scale mixing processes. As in Fig. \ref{fig:abundances_terminator_smallparticles} and \ref{fig:abundances_terminator_largeparticles}, the North pole is at the top and the morning terminator (leading limb) towards the right. The dashed white line indicates the pressure above which hazes are not allowed to exist.}
\label{fig:enhancedkzz_terminator}
\end{figure*}
We are first exploring how the 3D distribution of small-particle hazes changes with different values of the additional $K_{zz}$. Terminator cross-sections for a selection of $K_{zz}$ values are shown in Fig. \ref{fig:enhancedkzz_terminator}. For low $K_{zz}$ ($10^3$ m$^2$s$^{-1}$), the result is as expected very close to the simulations without additional sub-grid-scale mixing. For somewhat higher values ($10^4$ m$^2$s$^{-1}$,$10^5$ m$^2$s$^{-1}$, $0.01\times$Moses profile), the overall picture still looks relatively similar, however, there are less hazes in the center of the nightside vortices. Further, in the regions with a banded pattern (0.1-100 mbar), the bands with enhanced haze abundance do not move as much towards the equator with increasing pressure and the equatorial region is more depleted.

For $K_{zz}=10^6$ m$^2$s$^{-1}$ as well as the Moses profile, the diffusion timescale becomes comparable to or shorter than the vertical advection timescale even at high altitudes, where vertical velocities tend to be the strongest. For these simulations, the 3D distribution of hazes significantly changes. At low pressures ($p<10$ $\mu$bar), hazes are increasingly concentrated on the dayside and at high latitudes while large parts of the nightside, especially the nightside vortices, have low haze mixing ratios. This is because vertical mixing to deeper regions is faster than horizontal transport. The mass mixing ratio at the terminator drops by a factor of a few compared to the simulations without additional diffusion or with low eddy diffusion coefficients in that pressure region. In addition, the mass mixing ratio declines faster with increasing pressure. For $p>10$ $\mu$bar, the highest mass mixing ratios are found near the downwelling regions west of the antistellar point and at midlatitudes at the morning terminator. Overall, this picture results in relatively small differences between morning and evening terminator for $p<10$ $\mu$bar relative to the case of no sub-scale mixing  and larger mixing ratios at the morning terminator than at the evening terminator for $p>10$ $\mu$bar (see also Fig. \ref{fig:abundanceprofileplot_terminator}).

For even higher values of $K_{zz}$ ($10^7$ m$^2$s$^{-1}$), hazes are strongly concentrated on the dayside. Mixing ratios at the terminator are drastically reduced (by one to two orders of magnitude) compared to the other simulations. Hardly any hazes reach the nightside vortices. Because the nightside vortices still reach across the morning terminator, there are overall less hazes at the morning terminator than at the evening terminator.

\subsection{Transit spectra}
\begin{figure}
\includegraphics[width=\columnwidth]{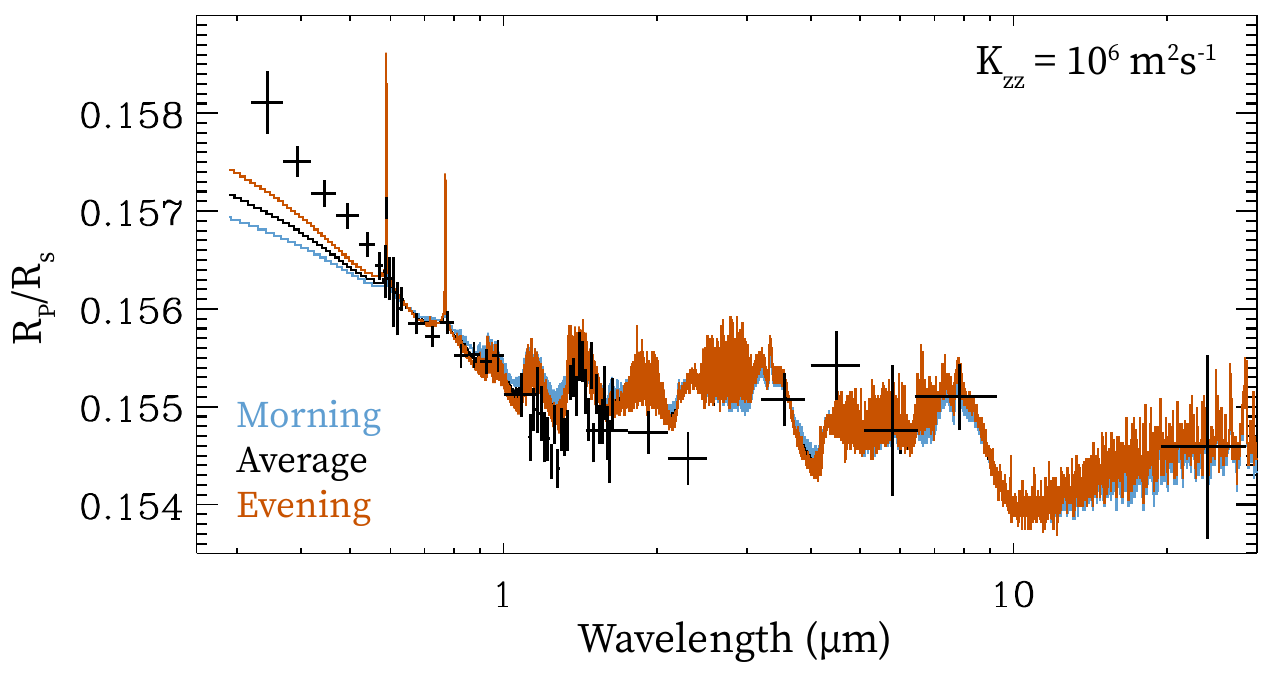}
\caption{Model transmission spectra for a high value of additional eddy diffusion and a particle radius of 3 nm, calculated for the morning (light blue) and evening (dark orange) terminator separately, as well as for the entire planet (black). The black crosses again represent observational data.}
\label{fig:spectra_3nm_enhancedkzz}
\end{figure}
For $K_{zz}<10^{6}$ m$^2$s$^{-1}$, the spectral slope remains relatively flat for all choices of the haze production rate. For the three simulations with high $K_{zz}$ ($10^{6}$ m$^2$s$^{-1}$, $10^{7}$ m$^2$s$^{-1}$, Moses profile), a much steeper slope arises, comparable to the observed slope. However, in the $10^{7}$ m$^2$s$^{-1}$ case, so few hazes reach the terminator that unrealistically high haze production rates are required to get a transmission spectrum that roughly matches the observations. We therefore choose to focus on the $10^{6}$ m$^2$s$^{-1}$ case to examine terminator differences in transmission. The spectrum for a particle size of 3 nm is shown in Fig. \ref{fig:spectra_3nm_enhancedkzz}.

The spectral slope at short wavelengths is steeper for the evening terminator. This is mainly because the mass mixing ratio decreases faster with increasing pressure at the evening terminator than at the morning terminator. This is mostly because the downwelling feature (hydraulic jump) near the morning terminator additionally concentrates hazes at midlatitudes at the morning terminator for pressures between 10 $\mu$bar and 1 mbar. The hotter temperature at the evening terminator additionally contributes to the steeper slope in the evening terminator spectrum.

At short wavelengths, the transit radius in Fig. \ref{fig:spectra_3nm_enhancedkzz} is larger for the evening terminator. However, this could be an artifact of the normalization procedure used, which assumes that the transit radius at 3.6 $\mu$m is the same for the morning and evening terminator. We therefore stress that the steeper slope at the evening terminator is a feature that arises directly from the 3D distribution of hazes in our model while the difference in the transit radius may depend on the normalization procedure for the transit spectrum.

\section{Discussion}
\label{sec:discussion}

\subsection{Limitations of GCM and haze model}
\label{subsec:disscussion_gcm}
The GCM simulations presented assume gray opacities. As discussed in Section \ref{subsec:radtran}, the gray model results in a qualitatively similar wind pattern compared to models with wavelength-dependent radiative transfer using the correlated-k method \citep[e.g.,][]{ShowmanEtAl2009,AmundsenEtAl2016}. We therefore expect that our key conclusions do not depend on the assumption of gray opacities. However, the peak velocities and the detailed location of some circulation features change when using more realistic radiative transfer. The detailed haze distribution may therefore differ from the results presented in this work when using the correlated-k method.

Two further key limitations of our GCM and our implementation of hazes are that our model neither includes radiative feedback from hazes nor haze growth. Hydrocarbon hazes are expected to be highly absorbing, especially if their optical properties are similar to soot \citep[e.g.,][]{MorleyEtAl2015,LavvasKoskinen2017}. Absorption and scattering of incoming radiation by high-altitude hazes can significantly affect the temperature profile \citep{MorleyEtAl2015,LavvasArfaux2021} and is likely to also change the atmospheric circulation, which in turn could lead to a 3D distribution of hazes differing substantially from the results presented in this study. This could alter the strength of vertical mixing, which potentially could lead to an improved fit to observations without the need to invoke additional sub-grid-scale mixing. We plan to address this topic in a future study.

Haze growth will also alter the 3D distribution of hazes. Based on 1D microphysics models \citep{LavvasKoskinen2017}, hazes are expected to be very small (1-3 nm) in the production region. Once they are mixed to regions with somewhat higher pressures ($>10^{-5}-10^{-4}$ bar), hazes can grow more efficiently through coagulation. Depending on the temperature profile, haze production rate and the efficiency of vertical mixing, haze particles can grow to sizes between 1 and 100 nm at 1 mbar and up to 1 $\mu$m at 1 bar. \citep{LavvasKoskinen2017}. In these 1D models, haze particles reach larger sizes when vertical transport is inefficient because there is more time for particles to grow through coagulation before they are transported to deeper layers. Combining the insights from these 1D studies of haze growth and our study, we can expect that when including haze growth in a 3D model, hazes would grow faster in regions where they are more concentrated such as within the nightside vortices for small particles. Because within our study, the 3D distribution is similar between different particle sizes within the small-particle-regime, one can expect that at low pressures, the mass mixing ratio distribution would remain similar but the particle size would vary horizontally. The 3D number densities would thus also deviate from our results. If particles in such a model (or a real atmosphere) are to grow to a size exceeding the small-particle regime, their distribution would likely differ significantly from the results presented for the large-particle regime in this work. The ``source regions'' for large particles would no longer be centered on the dayside but instead be the regions of largest growth of small particles, likely coinciding with regions of high concentrations of small particles. It will be important to examine these effects in future studies coupling a microphysics model to a GCM.

In regions of the atmosphere, in which particles grow to larger sizes, the shape of haze particles might also deviate from the spherical shape assumed in our simulation and instead form fractal aggregates. In the free-molecular-flow regime, the settling velocity of fractal aggregates with a fractal index of 2 is proportional to the monomer radius \citep{CabaneEtAl1993}. It can thus be expected that the settling velocity of fractal aggregates will be much smaller than that of spherical particles with the same mass. Therefore, the 3D mass mixing ratio distribution of aggregates at low pressures can be expected to be close to the one found in the small-particle regime. We point out that for the case of HD 189733b, 1D microphysical models \citep{LavvasKoskinen2017} predict haze particle sizes to be too small to form significant fractal aggregates at pressures probed in transit except in the case of weak vertical mixing throughout the atmosphere (including at low pressures, where there is strong vertical mixing in our simulations).

\subsection{Transit spectra}
\begin{figure}
\includegraphics[width=\columnwidth]{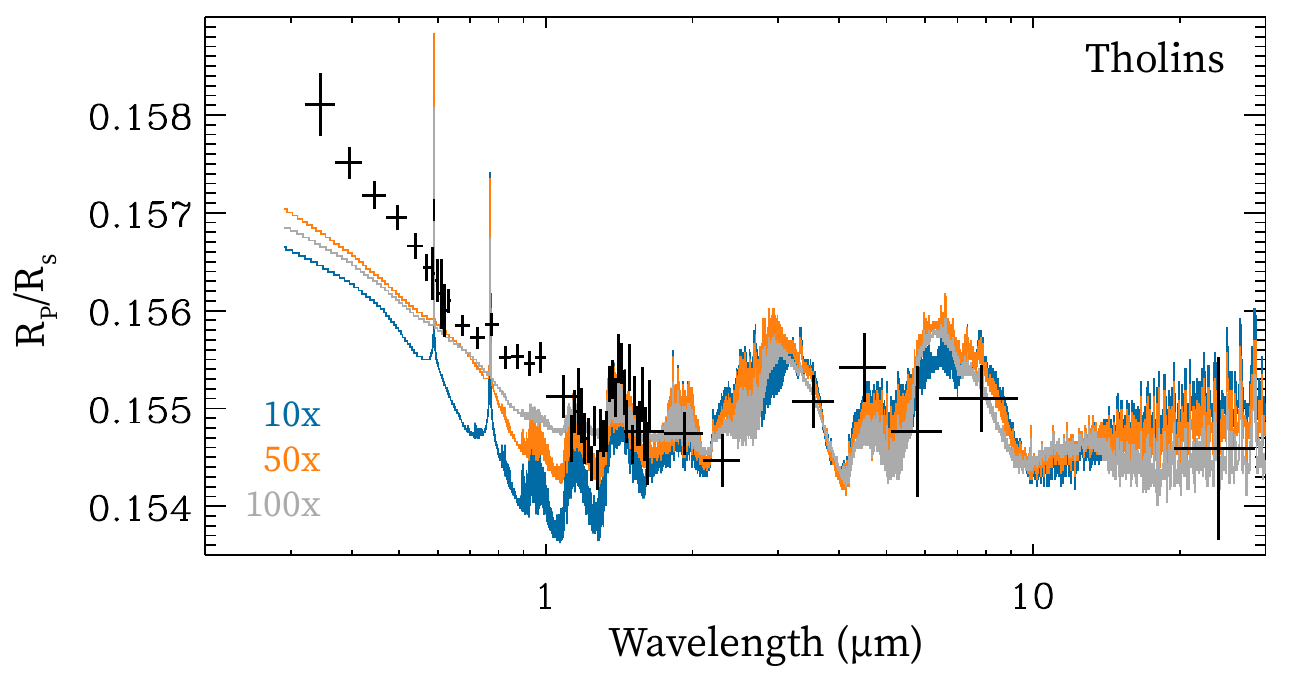}
\caption{Model transmission spectra using optical properties of tholins and a particle radius of 3 nm. Compared to the transmission spectrum calculated assuming a soot composition (Fig. \ref{fig:spectra_3nm}), the haze production rates have been increased by a factor of 10 (blue), 50 (orange) and 100 (light gray), corresponding to haze production rates at the substellar point of $2.5\cdot10^{-11}$ kg m$^2$s$^{-1}$, $1.25\cdot10^{-10}$ kg m$^2$s$^{-1}$ and $2.5\cdot10^{-10}$ kg m$^2$s$^{-1}$, respectively.}
\label{fig:tholins}
\end{figure}
There are multiple limitations of our model that affect the presented model transit spectra. We use a one-dimensional code with latitudinally averaged temperature and haze number density profiles to calculate transit spectra. Given the significant latitudinal variation of the haze mass mixing ratio, a two-dimensional \citep[e.g.,][]{MacDonaldMadhusudhan2017} or three-dimensional code \citep[e.g.,][]{FortneyEtAl2010,MillerRicciRauscher2012,CaldasEtAl2019,LeeEtAl2019MonteCarloRT} 
would be more appropriate. To address this concern, we conducted a test in which we split the morning terminator in multiple latitudinal segments, calculated the transit spectrum for the average mass mixing ratio profile within each segment and then combined the spectra for these segments. The same procedure was repeated for the evening terminator. Using more segments changed the transit spectra, mostly by slightly shifting the short-wavelength part of the spectra to  a lower transit radius, for a low number of segments. If more than six segments were used, the change was negligible. At the same time, the relative difference between the spectra of morning and evening terminator changed by much less. Thus, using 2D or 3D codes for calculating transmission spectra is more accurate and should be preferred in the future, but is unlikely to change our conclusions.

The radiative transfer within the GCM is double-gray and assumes opacities constant in pressure. This results in inaccurate temperatures at low pressures ($<0.1$ mbar). Typically, double-gray models overestimate the temperature in the pressure range between 0.1 mbar and 1 $\mu$bar compared to models with more sophisticated radiative transfer. Transit spectra are sensitive to the temperature profile and the uncertainty in the temperature profile could have a significant effect on the transit spectra. Nonetheless, our focus in this study has been on relative differences between morning and evening terminator. While simulations using more sophisticated radiative transfer, such as SPARC, find lower temperatures in the regions probed by transit, the temperature difference between morning and evening terminator remains similar to the temperature difference in the double-gray model. Therefore, we expect that including wavelength-dependent radiative transfer in the GCM would not change our main conclusions. Further, we briefly explored changing the temperature profile used in the transit spectrum calculation by using a temperature profile from a simulation using SPARC in the transit calculation. The change in the short-wavelength slope was relatively small. It certainly cannot account for the mismatch of the spectral slope between the model predictions from the nominal simulations and the observations. Matching the observed slope with the 3D distribution from our nominal simulations would require a temperature profile about three times hotter.

As mentioned earlier, radiative feedback from photochemical hazes could significantly heat the atmosphere at low pressures, potentially resulting in a steeper spectral slope. 1D models \citep[][]{MorleyEtAl2015} predict a temperature increase of up to a few hundred Kelvin. While this could somewhat lessen the mismatch between the model spectra and observations, it may not be enough to resolve the issue.

An additional limitation is that the optical properties of photochemical hazes are highly uncertain. While laboratory experiments have been able to produce haze analogs for temperature and chemical regimes expected for mini-Neptunes \citep{HorstEtAl2018,HeEtAl2018} and hot Jupiters \citep{FleuryEtAl2019,HeEtAl2020}, no  measurements of optical properties of these haze analogs have been reported yet. Similar to \citet{MorleyEtAl2013,MorleyEtAl2015} and \citet{LavvasKoskinen2017}, we assumed optical properties of soots.
\citet{OhnoKawashima2020} found that using tholin optical properties resulted in steeper short-wavelength slopes than for soots. A comparison of the optical properties of soots and tholins can be found in Fig. 2 of \citet{LavvasArfaux2021}. Tholins are not expected to be a good haze analog for hot Jupiters. They are produced in a nitrogen-dominated environment and at much lower temperatures. Nevertheless, we calculated transit spectra using absorption cross-sections of tholins as a test.  The resulting spectra indeed exhibited a steeper short-wavelength slope. We were able to roughly match the UV slope of the observed spectrum using the tholin optical properties and haze production rates 10 to 100 times higher than for the soot case (i.e. $2.5 \cdot 10^{-11}$ to $2.5 \cdot 10^{-10}$ kg m$^{-2}$s$^{-1}$). The resulting spectra are shown in Fig. \ref{fig:tholins}. We note that for tholins, the short-wavelength radius does not increase monotonically with increased haze opacity because the infrared spectral features of tholins affect the normalization of the spectrum. The better match of the short-wavelength slope should not be understood to mean that hazes on HD~189733b are similar to tholins in composition. It only means that the wavelength-dependence of their absorption cross-section might be closer to that of tholins than to that of soots. In other words, hazes with an absorption coefficient that steeply decreases from the NUV to the NIR could match observations more easily. Laboratory measurements show that the chemical composition of haze analogs is highly dependent on the temperature and composition of the initial gas mixture and that haze analogs for mini-Neptunes have incorporated much more oxygen than Titan haze analogs or soots \citep{MoranEtAl2020}. Therefore, real exoplanet haze particles could be very different from soots and from tholins. Measurements of the optical properties of the aerosols produced by these lab experiments will be crucial to inform the interpretation of observations.

In the case of hazes forming fractal aggregates, the optical properties would also differ. Compared to spherical particles of the same mass, fractal aggregates have a higher extinction cross-section in the UV but a lower extinction cross-section in the infrared \citep[e.g., Fig. 17 in][]{LavvasEtAl2011}. For a constant monomer radius, the larger the aggregate, the stronger is this effect. This could also increase the spectral slope in transit spectra. However, as pointed out earlier, it is relatively unlikely that large fractal aggregates form in the atmospheres of hot Jupiters.

The optical properties of the haze material might be further constrained by comparing model results to secondary eclipse observations. In particular, \citet{EvansEtAl2013} found a high value of the geometric albedo across 290-450 nm and a low value across 450-570 nm. This could indicate a strong wavelength dependence of the optical properties of the haze. We leave a detailed comparison of our model to secondary eclipse spectra across the full wavelength range of observations to future work that includes haze radiative feedback, which might strongly affect the infrared portion of the secondary eclipse spectrum.

\section{Conclusions}
\label{sec:conclusions}
We simulated the global distribution photochemical hazes in the atmosphere of hot Jupiter HD~189733b using a 3D general circulation model with gray radiative transfer and examined the implications for transmission spectroscopy. In our model, hazes are produced at low pressures (peak production: 2 $\mu$bar) on the dayside, are spherical with a constant particle size and do not exert any radiative feedback on the atmospheric circulation. Hazes are advected and can settle gravitationally.
We find that there are horizontal variations in the haze mass mixing ratio of at least an order of magnitude at all pressures within our simulation domain and for all particle sizes considered. The behavior of the hazes can be classified into two regimes: small particles (<30 nm) and large particles (>30 nm).

In the small-particle regime, gravitational settling is unimportant. The 3D distribution of hazes looks similar between different particle sizes within this regime. Near the peak of the production region (2 $\mu$bar), hazes are transported towards the poles and the nightside by the day-to-night flow prevalent at low pressures. At low pressures but below the peak haze production altitude (3 $\mu$bar $\lessapprox p\lessapprox0.1-1$ mbar), hazes accumulate in the two large vortices located at midlatitudes east of the anti-stellar point. Because these vortices extend across the morning terminator, small hazes are more abundant at the morning terminator than at the evening terminator at these pressures. The behavior of hazes in the small-particle regime thus is different from the predictions of \citet{KemptonEtAl2017}. Deeper in the atmosphere ($p\gtrapprox 1$ mbar), where the flow is dominated by the equatorial jet, hazes exhibit a more longitudinally symmetric, banded pattern and terminator differences are smaller.

In the large-particle regime, the 3D distribution strongly depends on the pressure at which the settling time scale and the horizontal transport time scale become comparable and the specific wind pattern at that pressure. However, some common patterns can be observed: Because of predominantly eastward winds on the dayside at the pressures where settling and horizontal transport time scales are comparable, there are more hazes at the evening terminator than at the morning terminator, consistent with the prediction of \citet{KemptonEtAl2017}. We note that, depending on the particle size, the peak mass mixing ratios at the terminator are, however, at mid- to high latitudes and not at the equator (as one might naively expect based on advection by the equatorial jet alone). At higher pressures ($>1$ mbar), where the equatorial jet is most efficient and horizontal advection is faster than settling, the haze distribution becomes more longitudinally symmetric, similar to the small particle regime.

We further derived an effective eddy diffusion coefficient $K_{zz}$ from the haze distribution in our simulations. Our suggested parametrization inferred from the simulations is $K_{zz} = 60 \cdot \left(\frac{P}{1 \textrm{bar}}\right)^{-0.9} $m$^2 $s$^{-1}$. This is a somewhat stronger pressure dependence than derived for cloud particles on HD~189733b in \citet{AgundezEtAl2014}. As a result, $K_{zz}$ is similar to the commonly used estimate $w_{\mathrm{rms}} \cdot H$ at very low pressures, near the haze production region, but much lower deeper in the atmosphere.

Examining the implications for transit observations, we focused on the small-particle regime, which is more consistent with observations and microphysics models. The difference between the transit spectrum of the morning terminator (leading limb in transit) and evening terminator is very small. We find that the effects of haze abundance differences and temperature differences between morning and evening terminator largely cancel each other out. We conclude that terminator differences due to hazes would be difficult to observe with low-resolution spectroscopy. Furthermore, even if such differences were to be observed, the interpretation could be ambiguous due to the opposing effects of temperature and haze abundance.

Furthermore, we were not able to match the observed steep short-wavelength slope with our nominal model. There are multiple factors that could explain this mismatch:
\begin{itemize}
    \item The optical properties of hazes present in the atmosphere of HD~189733b could significantly differ from those of soot. In particular, materials with an absorption cross-section that shows a stronger decrease from the near-UV to near-IR could better match the slope.
    \item Enhanced vertical mixing due to turbulent mixing at scales smaller than the grid size of the GCM could lead to a steeper slope that matches observations. Tentative simulations of that case show that a drastically different but still spatially inhomogeneous 3D haze distribution would be expected for that case. Our tentative simulations can better match the shape of the observed spectrum for sub-grid-scale mixing with a strength of $K_{zz}=10^6$ m$^2$s${-1}$.
    \item Hotter temperatures at low pressures due to heating by hazes absorbing starlight could result in a somewhat steeper slope, but this effect is likely not strong enough to resolve the mismatch between the model-predicted spectra and observations alone. 
    \item Radiative feedback from hazes could in addition change the atmospheric circulation and the rate of vertical mixing, which could result in a different 3D distribution of hazes and lead to a different short-wavelength slope.
    \item Finally, star spots could contribute to the observed short-wavelength slope of HD~189733b, as has been proposed previously \citep{McCulloughEtAl2014}.
\end{itemize}

Overall, our work demonstrates that the atmospheric circulation significantly shapes the 3D distribution of photochemical hazes in ways that cannot be captured by 1D models or post-processing of temperature profiles derived from 3D models. Follow-up studies to examine the interactions of atmospheric circulation with haze growth and radiative feedback of hazes will help to facilitate the interpretation of transmission spectra, secondary eclipse measurements and phase curves. We also stress the need to better constrain the optical properties of hazes that could be present in exoplanet atmospheres, for example by measurements of laboratory haze analogs.

\section*{Acknowledgements}
We thank Thaddeus Komacek for help with setting up the double-gray radiative transfer module and Yuan Lian for help with numerical issues with the MITgcm. This research was supported by NASA Headquarters under the NASA Earth and Space Science Fellowship Program - Grant 80NSSC18K1248. T.T.K acknowledges support by the NASA Exoplanet Research Program grant 80NSSC18K0569.
X. Z. acknowledges support from the NASA Solar System Workings Grant 80NSSC19K0791. This work benefited from the 2019 Exoplanet Summer Program in the Other Worlds Laboratory (OWL) at the University of California, Santa Cruz, a program funded by the Heising-Simons Foundation. We further thank Emily Rauscher for sharing the vertical velocities from her GCM for comparison. Special thanks goes to Vivien Parmentier and to the Oxford Exoplanets y'all Slack channel. M.S. would also like to thank Jess Vriesema for help with shell scripts, Herbert Steinrueck for conversations about hydraulic jumps and Robin Baeyens and Gilda Ballester for helpful discussions. Finally, we thank the reviewer for constructive comments that improved the manuscript.

This research made use of Numpy \citep{NumpyCitation}, SciPy \citep{SciPyCitation}, Matplotlib \citep{MatplotlibCitation} and NASA's Astrophysics Data System.

\section*{Data Availability}
The output of the 3D simulations presented in this article will be shared on reasonable request to the corresponding author.


\bibliographystyle{mnras}
\bibliography{biblio}



\appendix

\section{Changes to Difference Scheme in Gravitational Settling Term}
\label{sec:appendixsettlingscheme}
The gravitational settling term in the tracer equation (\ref{eq:tracereqn}) is given by $- g \partial F / \partial p$, where the downward flux due to gravitational settling has been abbreviated to $F=\rho \chi V_s$. In this section, the index $k$ is used for the vertical direction. For simplicity, the indices for the other two dimensions are omitted. To indicate that a quantity is defined at a cell center, the subscript $c$ is used (e.g., $p_{c,k}$, $\chi_{c,k}$). Quantities defined at the interface between cells (cell faces), are denoted with the subscript $f$ (e.g., $p_{f,k}$, $w_{f,k}$). The grid is set up such that the logarithm of the difference between the pressures at the cell faces $\log(\Delta p_f)$ remains constant. The grid points for scalar quantities (cell centers) are then centrally placed within these cells, i.e. $p_{c,k}=(p_{f,k}+p_{f,k+1})/2$. The index $k$ increases with decreasing pressure.

The previous version of the code, used by \citet{ParmentierEtAl2013} and \citet{KomacekEtAl2019}, used a central difference scheme:
\begin{equation}
\frac{\partial F}{\partial p} \biggr\rvert_{p=p_{c,k}} =- \frac{F_{f,k+1}-F_{f,k}}{\Delta p_f},   
\end{equation}
The minus sign accounts for the fact that the index k increases with decreasing pressure.

The updated difference scheme was adapted based on the scheme used in the model of \citet{LavvasEtAl2010} accounting for the fact that in our case, advection by resolved small-scale eddies replaces the role of diffusion. It was designed to smoothly transition from a central difference scheme to an upstream scheme depending on the magnitude of the settling velocity. For this purpose, we define the parameter $R=c V_{s;c,k}/w_{c,k}$ as the ratio between the settling velocity and the vertical velocity (in height-coordinates). Here, $c$ is an arbitrary parameter that controls the magnitude of the upstream correction. For the simulations presented in this publication, $c=1$. We further introduce the coefficients
\begin{eqnarray}
\alpha &=& \frac{\exp(R)}{\exp(R) + \exp(- R)}, \\
\beta &=& \frac{\exp(-R)}{\exp(R) + \exp(- R)}.
\end{eqnarray}

With these coefficients, the updated difference scheme can be written as
\begin{equation}
\frac{\partial F}{\partial p} \biggr\rvert_{p=p_{c,k}} = - \frac{2}{\Delta p_f} \left[ \alpha \left( F_{f,k+1} - F_{c,k}\right) + \beta \left( F_{c,k} - F_{f,k} \right) \right],
\end{equation}
where the minus sign again accounts for the direction of $k$. For $R \ll 1$, this scheme recovers the previously used central difference scheme. For $R \gg 1$, this scheme turns into an upstream scheme.


\bsp	
\label{lastpage}
\end{document}